\documentclass[epj]{svjour}

\usepackage{graphics}
\renewcommand{\Re}{\mathrm{Re\:}}
\renewcommand{\Im}{\mathrm{Im\:}}
\newcommand{\e}{\mathrm{e}}

\begin{document}
\title{Eikonal zeros in the momentum transfer space from
proton-proton scattering: An empirical analysis}
\author{R.F. \'Avila\inst{1} and M.J. Menon\inst{2}}

\mail{menon@ifi.unicamp.br}          % Insert a name or remove this line
\institute{Instituto de Matem\'atica, Estat\'{\i}stica e Computa\c c\~ao
Cient\'{\i}fica, Universidade Estadual de Campinas, 13083-970 Campinas, SP Brazil
\and
Instituto de F\'{\i}sica Gleb Wataghin,
Universidade Estadual de Campinas, 13083-970 Campinas, SP Brazil}
\date{Received: date / }

\abstract{
By means of improved empirical fits to the differential cross section
data on $pp$ elastic scattering at $19.4 \le\sqrt{s}\le 62.5$
GeV and making use of a semi-analytical method,
we determine the eikonal in the momentum transfer space
(the inverse scattering problem). This method allows 
for the propagation of
the uncertainties from the fit parameters up
to the extracted eikonal, providing statistical evidence
that the imaginary part of the eikonal (real part of the
opacity function) presents a zero (change of signal) in
the momentum space, at $q^2 \approx 7 \pm 1$ GeV$^2$. We
discuss the implication of this change of signal in the
phenomenological context, showing that eikonal models with one zero
provide good descriptions of the differential cross sections in the
full momentum transfer range, but that is not the case for
models without zero. Empirical connections between the extracted
eikonal and results from a recent global analysis on the proton electric form factor are also discussed,
in particular the Wu-Yang conjecture.
In addition, we present a
critical review on the $pp$ differential cross section data
presently available at high energies.
 % end of PACS codes
} %end of abstract
\authorrunning{R.F. \'Avila and M.J. Menon }

\titlerunning{Eikonal zeros in the momentum transfer space
from proton-proton scattering}

\maketitle

\noindent
\textbf{Contents}

\vspace{0.3cm}

\begin{description}

\item[1] Introduction

\vspace{0.3cm}

\item[2] Eikonal representation and the inverse scattering problem

\vspace{0.3cm}

\item[3] Critical discussion on $pp$ differential cross section data
at high energies

\vspace{0.1cm}

\subitem 3.1 Data at 23.5 $\leq \sqrt s \leq$ 62.5 GeV
 
\vspace{0.1cm}

\subitem 3.2 Data at 19.4 and 27.4 GeV

\vspace{0.1cm}

\subsubitem $3.2.1$ $\sqrt s$ = 19.4 GeV

\vspace{0.1cm}

\subsubitem $3.2.2$ $\sqrt s$ = 27.4 GeV

\vspace{0.1cm}

\subitem 3.3 Discussion on data at large momentum transfers

\vspace{0.1cm}

\subsubitem $3.3.1$ References and data

\vspace{0.1cm}

\subsubitem $3.3.2$ Dependence on the energy

\vspace{0.3cm}

\item[4] Improvements in the previous analysis and fit results

\vspace{0.1cm}

\subitem 4.1 Data ensembles and energy independence

\vspace{0.1cm}

\subitem 4.2 Parametrization

\vspace{0.1cm}

\subitem 4.3 Confidence intervals for uncertainties

\vspace{0.1cm}

\subitem 4.4 Fit results

\vspace{0.3cm}

\item[5] Eikonal in the momentum transfer space

\vspace{0.1cm}

\subitem 5.1 Semi-analytical method

\vspace{0.1cm}

\subitem 5.2 Eikonal zeros

\vspace{0.3cm}

\item[6] Phenomenological implication of the eikonal zero

\vspace{0.1cm}

\subitem 6.1 Eikonal and scattering amplitude

\vspace{0.1cm}

\subitem 6.2 Some representative eikonal models

\subsubitem $6.2.1$ Models without eikonal zero

\ \ \ \ \ \ \ \ \ \  $\bullet$ Chou-Yang model

\ \ \ \ \ \ \ \ \ \ $\bullet$ QCD-inspired models

\subsubitem $6.2.2$ Hybrid model

\subsubitem  $6.2.3$ Models with eikonal zero

\ \ \ \ \ \ \ \ \ \ $\bullet$ Bourrely-Soffer-Wu model

\ \ \ \ \ \ \ \ \ \ $\bullet$ A multiple diffraction model

\subsubitem  $6.2.4$ Discussion

\ \ \ \ \ \ \ \ \ \   $\bullet$ General aspects

\ \ \ \ \ \ \ \ \ \ $\bullet$ Quantitative aspects

\ \ \ \ \ \ \ \ \ \ $\bullet$ Empirical parametrization for the eikonal

\vspace{0.1cm}

\subitem 6.3 Hadronic and electromagnetic form factors

\subsubitem $6.3.1$ Rosenbluth and polarization transfer results

\subsubitem  $6.3.2$ The Proton electric form factor

\vspace{0.3cm}

\item[7] Summary and final remarks

\end{description}

%\newpage

\section{Introduction}
\label{sec:1}

Quantum Chromodynamics (QCD) is very successful in describing 
hadronic scattering involving very large momentum transfers
\cite{gelis}. However, that is not the case for \textit{soft diffractive 
processes} (large distance phenomena), in particular the simplest process:
high-energy elastic hadron scattering. The point is that perturbative
techniques can not be applied and presently, non-perturbative
approaches can not describe scattering states without strong
model assumptions \cite{pred,donna}. 
At this stage phenomenology is an important approach and 
among the wide variety of models, the eikonal picture plays a central role
due to its connection with unitarity \cite{pred}.

Alongside phenomenological models, \textit{empirical analyses}, aimed to extract
\textit{model-independent information} from the experimental data (the inverse
scattering problem), also constitute important strategy that can contribute with
the establishment of novel theoretical calculational schemes. In an 
unitarized context this approach is characterized by model-independent
extraction of the eikonal from empirical fits to the differential cross
section data, mainly on proton-proton ($pp$) and antiproton-proton ($\bar{p}p$)
scattering (highest energies reached in accelerator experiments).
However, one of the main problems with this kind of analysis
is the very limited interval of the momentum transfer with available data,
in general below 6 GeV$^2$. This means that, from the statistical point
of view, all the extrapolated curves from the fits must be taken into
account, which introduces large uncertainties in the extracted
information. 

In references \cite{cmm,cm} this problem was addressed through a detailed analysis
of the experimental data in the region of large momentum transfer and that allowed
the extraction to be made of the eikonal on statistical grounds. The main result,
from the analysis of  $pp$ elastic scattering at $19.4 \le\sqrt{s}\le 62.5$, 
concerned the
evidence of eikonal zeros (change of sign) in the momentum transfer space and that
the position of the zero decreases as the energy increases \cite{cmm}.
As discussed in that paper, this kind of model independent information in 
the momentum space is very important in the construction and selection of
phenomenological approaches, mainly in the case of diffraction models,
since the eikonal, in the momentum transfer space,
is expected to be connected with hadronic form factors and elementary
cross sections.

In this work, we introduce two main improvements in this previous analysis,
which are related with the ensemble of the selected data and the structure of
the parametrization. We still obtain statistical evidence for the zeros, but 
different from \cite{cmm}, it can not be inferred that the position of the 
zero decreases as the energy increases (some preliminary results on this 
feature
appear in \cite{amm}).
In order to explain some subtleties involved in the analysis,
we present a novel critical review and discussions on the $pp$ differential cross 
section data presently available. That may be very opportune since 
presently a great development is expected of the area with the next
$pp$ experiments at 200 GeV (BNL RHIC) \cite{pp2pp} and 14 TeV (CERN LHC) 
\cite{totem}. In addition, we discuss in some detail the implication of the 
eikonal zero
in the phenomenological context, introducting a novel analytical
parametrization for the extracted eikonal. Connections between the empirical
result for the eikonal and recent data on the proton electric form
factor are also presented and discussed, in particular the Wu-Yang conjecture. 

The manuscript is organized as follows.
In Sect.~\ref{sec:2} we recall the main formulas
connecting the experimental data and the eikonal (the inverse scattering problem).
In Sect.~\ref{sec:3} we present a critical review on the experimental
data pres\-ent\-ly available from elastic $pp$ scattering.
In Sect.~\ref{sec:4} we discuss
the improvements introduced in the previous analysis and present the new fit
results.  
In Sect.~\ref{sec:5} we
treat the determination of the eikonal in the momentum transfer space
and in Sect.~\ref{sec:6} we discuss the implication of the eikonal zeros in the 
phenomenological context, as well as connections between the extracted eikonal, models 
and the proton electric form factor. 
The conclusions and some final remarks are the contents of Sect.~\ref{sec:7}.

\section{Eikonal representation and the inverse scattering problem}
\label{sec:2}

In the eikonal representation,
the elastic scattering amplitude can be expressed by
\cite{pred}

\begin{equation}
F(s, q) = \mathrm{i} \int_{0}^{\infty} b\mathrm{d}b J_{0}(qb)
\{ 1 - \e^{\mathrm{i}\chi(s,b)}\},
\label{eq:1}
\end{equation}
where $s$ is the center-of-mass energy squared, $q^2=-t$ the four-momentum transfer 
squared, $b$ the impact parameter and $\chi(s,b)$ the eikonal
function in the impact parameter space (azimuthal symmetry assumed). 
It is also useful to define the
Profile function (the inverse transform of the amplitude) in terms
of the eikonal:

\begin{equation}
\Gamma(s,b)=
 1 - \e^{\mathrm{i}\chi(s,b)}. 
\label{eq:2}
\end{equation}
With these definitions the complex eikonal corresponds to the continuum
complex phase shift, in the limit of high energies and the semi-classical
approximation: $\chi(s,b) = 2\delta(s,b)$; that is also the normalization 
in the Fraunhofer regime \cite{pred}.

In the theoretical context, 
eikonal models are characterized by different phenomenological choices for the 
eikonal
function in the \textit{momentum transfer space}:

\begin{equation}
\tilde\chi(s,q) =  \int_{0}^{\infty} b \mathrm{d}b J_{0}(qb) \chi(s,b).
\label{eq:3}
\end{equation}

On the other hand, the \textit{inverse scattering problem} deals with the empirical 
determination,
or extraction, of the eikonal from the experimental data on
the differential cross section

\begin{equation}
\frac{\mathrm{d}\sigma}{\mathrm{d}q^2} = \pi|F(s, q)|^2,
\label{eq:4}
\end{equation}
the total cross section (optical theorem)

\begin{equation}
\sigma_{\mathrm{tot}}(s) = 4\pi \mathrm{Im}\ F(s, q=0),
\label{eq:5}
\end{equation}
and the parameter $\rho$, defined as the ratio of the real to
the imaginary part of the forward amplitude,

\begin{equation}
\rho(s) = \frac{\mathrm{Re}\ F(s, q=0)}{\mathrm{Im}\ F(s, q=0)}.
\label{eq:6}
\end{equation}

Formally, from a model independent parametrization for the 
scattering amplitude
and fits to the differential cross section data, one can extract the profile function

\begin{equation}
\Gamma(s,b)=-\mathrm{i}\int_0^{\infty} q \mathrm{d} q J_0(qb)F(s,q),
\label{eq:7}
\end{equation}
the eikonal in the impact parameter space,

\begin{equation}
\chi(s,b)=-\mathrm{i}\ln\left[1-\Gamma(s,b)\right],
\label{eq:8}
\end{equation}
and then, \textit{under some conditions}, the eikonal in the momentum transfer space 
through Eq.~(\ref{eq:3}). The possibility to extract $\tilde\chi(s,q)$ is very important if we 
look for possible connections with quantum field theory since elementary
(partonic) cross sections
are expressed in the momentum transfer space as well as form factors
of the nucleons.

However, as already commented on in our introduction, we stress
that a drawback of the above 
\textit{inverse scattering}
is the fact that the differential cross section data available cover 
only limited 
regions in terms of the momentum transfer, which in general is small,
as referred and discussed in what follows. This can be contrasted with the fact that, 
in order
to extract the eikonal,
all the Fourier-Bessel transforms must be performed in the interval
$0 \rightarrow \infty $. Therefore, \textit{data at large values of the momentum
transfer} play a central role in this kind of analysis and for that reason
we first discuss in the next section the experimental data presently 
available and some subtleties involved in the selection, normalization
and interpretation of the data sets. A detailed analysis of data at small
$q^2$ is discussed in \cite{cudell} and extended in \cite{clm}.

\section{Critical discussion on $pp$ differential cross section data  
at high energies}
\label{sec:3}

In Ref. \cite{cmm} it has been shown that the
lack of sufficient experimental information on $\bar{p}p$
elastic differential cross sections does not allow to perform the kind of analysis we
are interested in. For that reason we shall treat here only
$pp$ elastic scattering at the highest energies, namely $\sqrt s$
above $\approx $ 19 GeV. 

The inputs of our analysis concern the experimental data on
differential cross section, total cross section, the
$\rho$ parameter and the corresponding optical point:

\begin{equation}
\left.\frac{\mathrm{d}\sigma}{\mathrm{d}q^2} \right|_{q^2=0} =
\frac{\sigma_{\mathrm{tot}}^2 (1 + \rho^2)}{16\pi}, 
\label{eq:9}
\end{equation}
where $\rho(s)$ and $\sigma_{\mathrm{tot}}(s)$ are the experimental values
at each energy. 
Since we are interested only in the hadronic interaction, the selected differential
cross section data cover the region above the Coulomb-nuclear interference,
namely $q^2 >$ 0.01 GeV$^2$. 
In what follows we discuss the sets of data at 7 different energies, divided
in the two groups 23.5 $\leq \sqrt s \leq$ 62.5 GeV  (5 sets) and $\sqrt s$ =
19.4 and 27.4 GeV.

\subsection{Data at 23.5 $\leq \sqrt s \leq$ 62.5 GeV}
\label{sec:3.1}

The five data sets at $\sqrt s$ = 23.5, 30.7, 44.7, 52.8 and 62.5 GeV
were obtained at the CERN Intersecting Storage Ring (ISR) in
the seventies and still represent the largest and highest energy 
range of available data on $pp$ scattering (the recent experiment
at RHIC by the $pp2pp$ Collaboration measured only the slope
parameter at $\sqrt s$ = 200 GeV \cite{pp2ppslope}).
The data on $\sigma_{\mathrm{tot}}$, $\rho$ and 
$\mathrm{d}\sigma/\mathrm{d}q^2$ were  
compiled and analyzed by Amaldi and Schubert leading to the most 
coherent set of data on $pp$ scattering. Detailed information on this analysis
can be found in \cite{as}; here we only recall some relevant aspects to
our discussion.

\vspace{0.3cm}

\noindent
\textit{Optical points}.
The numerical values for the total cross sections, correspond to the
average of 3 experiments, performed in each of the above energies;
the $\rho$ data comes from 2 experiments, one at 23.5 GeV and 
the other in the region 30.7 - 62.5 GeV.
These numerical values are displayed in Table~\ref{tab:1}, together with the
corresponding optical points (and references for the $\rho$ data).

\begin{table*}
\begin{center}
\caption{Forward data ($\sigma_{\mathrm{tot}}$ and $\rho$) and optical points from 
$pp$ scattering used in this analysis.}
\label{tab:1}
\begin{tabular}{cccc}
\hline
$\sqrt{s}$  & $\sigma_{\mathrm{tot}}$ & $\rho$ & $\mathrm{d}\sigma/\mathrm{d}q^2|_{q^2=0}$ \\
  (GeV)     &   (mb)        &        &  (mbGeV$^{-2}$) \\
\hline
19.4 & 38.98  $\pm$ 0.04 \cite{car} & 0.019  $\pm$ 0.016 \cite{faj} & 77.66  $\pm$ 0.02    \\
23.5 & 38.94 $\pm$ 0.17 \cite{as} & 0.02 $\pm$ 0.05 \cite{a1} & 77.5 $\pm$ 0.7  \cite{as} \\
30.7 & 40.14 $\pm$ 0.17 \cite{as} & 0.042 $\pm$ 0.011 \cite{a2} &82.5 $\pm$ 0.7 \cite{as} \\
44.7 &41.79 $\pm$ 0.16 \cite{as} &0.0620 $\pm$ 0.011 \cite{a2} &89.6 $\pm$ 0.7  \cite{as}\\
52.8 &42.67 $\pm$ 0.19 \cite{as} &0.078 $\pm$ 0.010 \cite{a2}& 93.6 $\pm$ 0.8  \cite{as}\\
62.5 &43.32 $\pm$ 0.23 \cite{as} &0.095 $\pm$ 0.011 \cite{a2}&96.8 $\pm$ 1.1  \cite{as} \\ 
\hline
\end{tabular}
\end{center}
\end{table*}

\vspace{0.3cm}

\begin{figure}
\resizebox{0.48\textwidth}{!}{\includegraphics{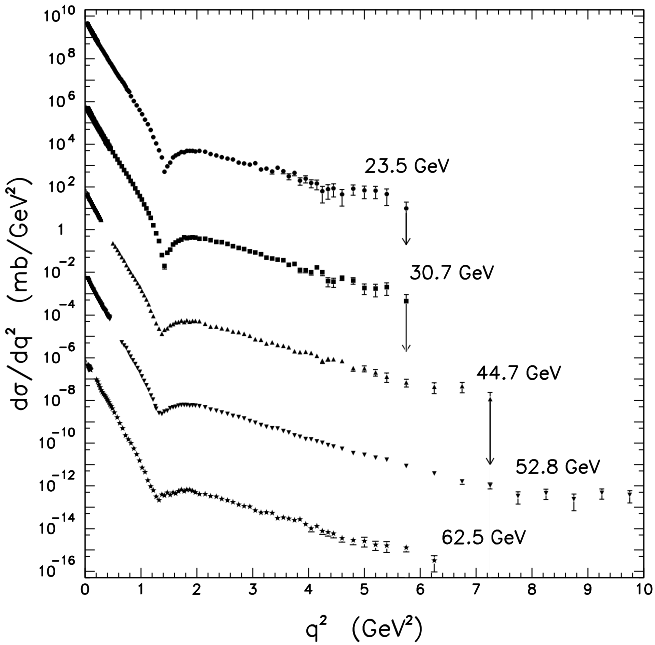}}
\caption{Proton-proton differential cross section data at the ISR energy
region from Landolt B\"ornstein tables \cite{lb} and optical points
from Table~\ref{tab:1}. Data were multiplied by factors of $10^{\pm 4}$.}
\label{fig:1} 
\end{figure}

\noindent
\textit{Data beyond the forward direction} ($q^2 > $ 0.01 GeV$^2$). 
The data in the region of small momentum transfer were normalized to
the optical point and above this region different data sets were normalized 
relative to each other, taking into account both the statistical and 
systematic errors (see \cite{as} for details). The final result of this
coherent and accurate analysis of the differential cross sections were
published in the numerical tables of the series Landolt-B\"ornstein (LB)
\cite{lb}, from which we extracted our data sets. They are reproduced in 
Fig.~\ref{fig:1},
together with the optical points (Table~\ref{tab:1}). As we can see, the largest set 
with available data correspond to $\sqrt s$ = 52.8 GeV, with
$q_{\mathrm{max}}^2$ = 9.75 GeV$^2$. Except for the data at 44.7 GeV all the
other sets cover the region nearly below 6 GeV$^2$.

\subsection{Data at $\sqrt s$ = 19.4 and 27.4 GeV}
\label{sec:3.2}

These sets correspond to the largest values of the momentum transfer with 
available data, namely $q_{\mathrm{max}}^2$ = 11.9 GeV$^2$ ($\sqrt s$ = 19.4 GeV) and
$q_{\mathrm{max}}^2$ = 14.2 GeV$^2$ ($\sqrt s$ = 27.4 GeV). For that reason they play 
a fundamental role in our analysis. In especial, as we shall discuss,
data at 27.4 GeV are crucial for the statistical evidence of the eikonal
zero and data at 19.4 GeV is extremely important for giving information
on the energy dependence of the position of the zero.

These data were obtained in the seventies-eighties at the 
Fermi National Accelerator Laboratory (Fermilab)
and at the CERN Super Proton Synchrotron (SPS). 
Some of these data sets were then published or made available from the authors
in a preliminary form. We call attention to this fact because
comparisons and interpretation may occur
that are not consistent with what can be inferred from the final
published results. In this section we first list and summarize our
selection of the experimental data and then discuss the information that can be 
extracted from these ensembles.

\subsubsection{$\sqrt s$ = 19.4 GeV}
\label{sec:3.2.1}

\noindent
\textit{Optical point}. We evaluate the optical point, Eq.~(\ref{eq:9}), with the values 
of the total
cross section obtained by Carrol et al.
\cite{car} and the $\rho$ parameter  by Fajardo et al. \cite{faj}
(Table~\ref{tab:1}).
Both experiments were performed at the Fermilab with 
beam momentum $p_{\mathrm{lab}}$ = 200 GeV ($\sqrt s$ = 19.42 GeV).

\vspace{0.3cm}

\noindent
\textit{Data beyond the forward direction}. We made use of the
following data sets:

\vspace{0.2cm}

\noindent
0.075 $\leq q^2 \leq$ 3.25 GeV$^2$.
Final data published by Akerlof et al. \cite{akerlof} and obtained
at the Fermilab with $p_{\mathrm{lab}}$ = 200 GeV. The errors are statistical and 
the absolute
normalization uncertainty is 7 \%.

\vspace{0.2cm}

\noindent
5.0 $\leq q^2 \leq$ 11.9 GeV$^2$.
Final results from Faissler et al. \cite{faissler}, obtained
at the Fermilab with $p_{\mathrm{lab}}$ = 201 GeV
($\sqrt s$ = 19.47 GeV). The errors are statistical and 
overall normalization error is 15 \%.

\vspace{0.2cm}

\noindent
0.6125 $\leq q^2 \leq$ 3.90 GeV$^2$.
Data obtained by Fidecaro et al. \cite{fidecaro}, 
at the CERN-SPS with $p_{\mathrm{lab}}$ = 200 GeV.
The $q^2$ values correspond to the central values of the bins from
[0.600 - 0.625] to [3.8 - 4.0].
The data are normalized \cite{fidecaro} and the errors are statistical.

\vspace{0.2cm}

\noindent
0.95 $\leq q^2 \leq$ 8.15 GeV$^2$.
Final results from Rubinstein et al. \cite{rubinstein}, obtained
at the Fermilab with $p_{\mathrm{lab}}$ = 200 GeV.
The errors are statistical and systematic uncertainties in overall 
normalization are 15 \%. The points at $q^2$ = 6.55 and 8.15 GeV$^2$ have
a statistical error of 100 \%.

\subsubsection{$\sqrt s = 27.4$ GeV}

The data cover the region
5.5 $\leq q^2 \leq$ 14.2 GeV$^2$ and we have used the
final results from Faissler et al. \cite{faissler}, obtained
at the Fermilab with $p_{\mathrm{lab}}$ = 400 GeV
($\sqrt s$ = 27.45 GeV). The errors are statistical and 
overall normalization error is 15 \%.

\vspace{0.3cm}

\begin{figure}
\resizebox{0.48\textwidth}{!}{\includegraphics{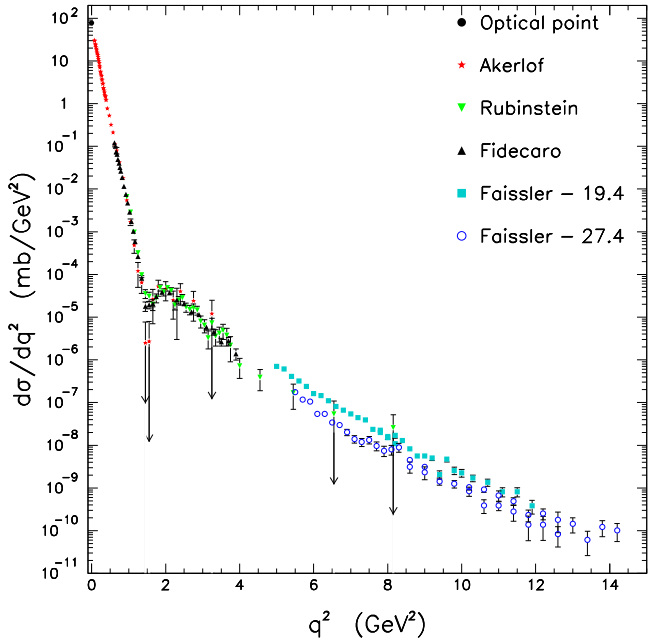}}
\caption{Differential cross section data at $\sqrt s$ = 19.4 GeV
and 27.4 GeV used in this analysis.}
\label{fig:2}       
\end{figure}

All these data at 19.4 and 27.4 GeV are displayed in Fig.~\ref{fig:2} with the
statistical errors. The intervals in the momentum transfer of all
data referred to above, $19.4 \le\sqrt{s}\le 62.5$
GeV, are summarized in Table~\ref{tab:2}.

\begin{table*}
\begin{center}
\caption{Intervals in the momentum transfer for the
differential cross section data
at $q^2>0.01$ GeV$^2$ (above the Coulomb-nuclear
interference region), number of points used in this analysis
and references for the tables.}
\label{tab:2}
\begin{tabular}{cccc}
\hline
$\sqrt{s}$ (GeV)  & $q^2$ interval (GeV$^2$) & Number of points & References\\
\hline
19.4 (Fermilab and CERN-SPS) & \ 0.075 - 11.9 &  156 & 
\cite{akerlof,faissler,fidecaro,rubinstein} \\
23.5 (CERN-ISR)              & \ 0.042 - 5.75 & 172 & \cite{lb} \\
27.4 (Fermilab)              & \ 5.5 - 14.2 & 39 & \cite{faissler} \\
30.7 (CERN-ISR)              & \ 0.016 - 5.75 & 211 & \cite{lb} \\
44.7 (CERN-ISR)              & \ 0.01026 - 7.25 & 246 & \cite{lb} \\
52.8 (CERN-ISR)              & \ 0.01058 - 9.75 & 244 & \cite{lb} \\
62.5 (CERN-ISR)              & \ 0.01074 - 6.25 & 163 & \cite{lb} \\
\hline
\end{tabular}
\end{center}
\end{table*}

\subsection{Discussion on data at large momentum transfers}

We now focus the discussion on the experimental data available
in the region of large momentum transfer, which, as already
noted, play a central role in the global information that can be
extracted from the fit procedure (uncertainty region and error 
propagation). We first call attention to some differences appearing
in the published data and then discuss the dependence on the energy
of data above $q^2$ = 3 - 4 GeV$^2$ in the region of interest
19 - 63 GeV (Figures \ref{fig:1} and \ref{fig:2}). To clarify some points we shall
follow a nearly chronological order.

\subsubsection{References and data}
\label{sec:3.3.1}

As we have seen, the \textit{final data} from the CERN-ISR
at 23.5 $\leq \sqrt s \leq$ 62.5 GeV, compiled and normalized
by Amaldi and Schubert, were published in LB tables in 1980.
Concerning this ensemble it should be noted that
final data from experiments, at large momentum transfer,
were previously published by Nagy et al. in 1979 \cite{nagy},
covering the region above 0.825 GeV$^2$ ($\sqrt s$ = 23.5, 52.8. 62.5)
and above 0.975 GeV$^2$ ($\sqrt s$ = 30.7, 44.7). The point here is
that although the authors refer to final results, the numerical
values appearing in the LB tables are about 3 \% higher than those
by Nagy et al.. This difference may be due to the normalization process by
Amaldi and Schubert, referred to in Sect.~\ref{sec:3.1}.

The data at 19.4 and 27.4 also appear in the LB tables and in this
case we first note that: 1) these data did not take part in the analysis by
Amaldi and Schubert (only ISR data) being, therefore, not normalized;
2) some numerical values appearing in the tables are
preliminary and do not correspond to final published results, as discussed
in what follows; 3) other data at 19.4 were published after 1980
\cite{fidecaro,rubinstein} (Sect. \ref{sec:3.2}).

\begin{figure}
\resizebox{0.48\textwidth}{!}{\includegraphics{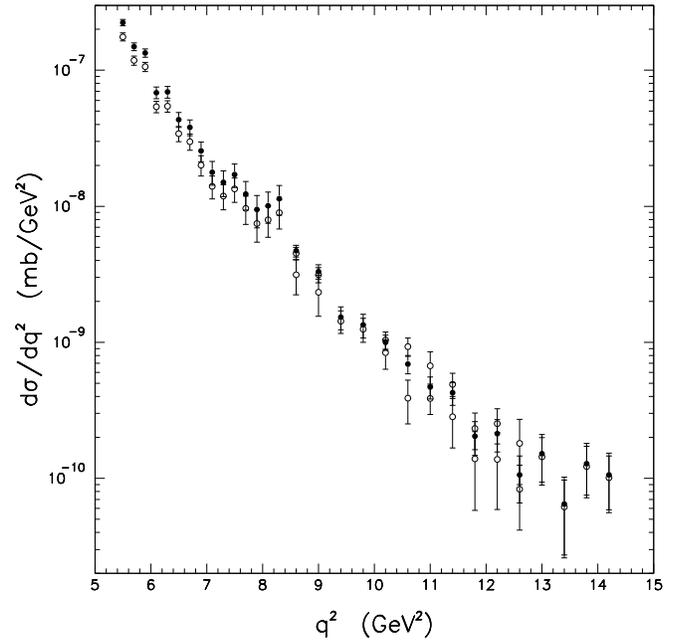}}
\caption{Differential cross section data on $pp$ scattering
at 27.4 GeV from the Landolt-B\"ornstein
tables \cite{lb} (black circles) and from Faissler et al.
\cite{faissler} (white circles). The uncertainties correspond
to the statistical errors only.}
\label{fig:3}       
\end{figure}

At 19.4 GeV, the data appearing in the LB tables in the region
0.075 $\leq q^2 \leq$ 3.25 GeV$^2$ are exactly the same as those
published by Akerlof et al. in 1976 \cite{akerlof}. However,
data at this energy in the region 5.5 $\leq q^2 \leq$ 11.9 GeV$^2$
and those at 27.4 GeV and 5.5 $\leq q^2 \leq$ 14.2 GeV$^2$
do not correspond to the final values published by Faissler et al.
in 1981 \cite{faissler}. The differences, in the case of data at
27.4 GeV, are illustrated in Fig.~\ref{fig:3}, where we see that although the general
trend of both sets are similar, the corrections are different in different
regions of the momentum transfer (the geometries of the experiment
at mid and high $q^2$ values \cite{faissler}). Moreover, the preliminary
set appearing in the LB tables has 30 data points and the final set by
Faissler et al. presents 39 data. We shall return to this point
when discussing the improvements in our previous analysis (Sect. \ref{sec:4.1}).

\subsubsection{Dependence on the energy} 

It has been argued that the data on $pp$ scattering at large momentum
transfers ($q^2 > $ 3 - 4 GeV$^2$) and energies above
$\sqrt s \sim$ 19 GeV have a small dependence on the energy.
That represents an important aspect because, if this dependence can
be neglected, the information at the largest values of the momentum
transfer (for example, data at 27.4 GeV) can be added to sets at nearby
energies leading to a drastic reduction of the uncertainty
regions in fit procedures. In fact that was the strategy in previous
analysis that allowed the statistical evidence 
to be inferred for the eikonal
zeros \cite{cmm,cm}. However, the exact value of the energy and momentum
transfer above which this dependence can be neglected is not clear
in the literature. In what follows, we first recall some previous
results, comparisons and arguments and then present a quantitative
test which allow to infer numerical limits or bounds for the
independence on the energy. 

In this respect the main ensemble is obviously the data at 19.4 and 27.4,
published by Faissler et al. in 1981, since they cover the region
up to 11.9 and 14.2 GeV$^2$, respectively. One important result
of this measurement was that the data showed no sign of a second dip at large
momentum transfer (this dip was previously suggested by the ISR
data at 52.8 GeV and $q^2 \sim$ 8 - 9 GeV$^2$ - Fig.~\ref{fig:1}). Faissler et al.
indicate that
the ratio of the differential
cross section data at 19.4 and 27.4 GeV, for the same $q^2$ and interval analyzed, is
about 2.3. This difference can be seen in Fig.~\ref{fig:2}, indicating therefore
a reasonable energy dependence. The authors also present comparison of the
data at 27.4 GeV and those at 52.8 GeV (ISR). For the ISR data they
quote the paper published by De Kerret et al. in 1977 \cite{dekerret},
where no table is available, only a plot of the data; there is also
no reference to the \textit{final} values published by Nagy et al. 
in 1979 \cite{nagy}.
According to Faissler et al.,  comparison of data at 27.4 GeV with those preliminary results at
52.8 GeV indicated a ratio of 1.5 $\pm$ 0.3, after taking into account 
the normalization errors quoted in both experiments.
The authors conclude that the energy dependence is significantly less for
$\sqrt s > $ 27.4 GeV than it is for $\sqrt s < $ 27.4 GeV, referring to a small
energy dependence beyond 27.4 GeV for 5 $ < q^2 < $ 8 GeV$^2$ \cite{faissler}.

Another aspect discussed by these authors concerns the 
\textit{value of the
slope} of the differential cross section at large momentum transfers. In particular,
they show that the data at 27.4 GeV follow a power fit of the form
$(q^2)^{-\lambda}$
with $\lambda =$ 8.45 $\pm$ 0.1 and $\chi^2$ = 33 for 28 degrees of freedom
\cite{faissler}.
This result was interpreted as consistent with the QCD multiple-gluon-exchange
calculation by Donnachie and Landshoff, which predicts
$\lambda$ = 8 \cite{land}.

In order to get some quantitative and detailed information
on the energy dependence of the data at large momentum transfer,
$q^2 >$ 3 - 4 GeV$^2$, and in the energy region of interest
(19 - 63 GeV), we have performed several tests with our
selected data (Sect. \ref{sec:3.1} and \ref{sec:3.2}), taking into account only the
statistical errors. We consider the following parametrization

\begin{equation}
\frac{\mathrm{d}\sigma}{\mathrm{d} q^2} = \frac{K}{(q^{2}/Q^{2})^{\lambda}},
\label{eq:10}
\end{equation}
with $Q^{2}$ = 1 GeV$^2$, so that $K$ is given in mbGeV$^{-2}$.

\begin{figure}
\resizebox{0.48\textwidth}{!}{\includegraphics{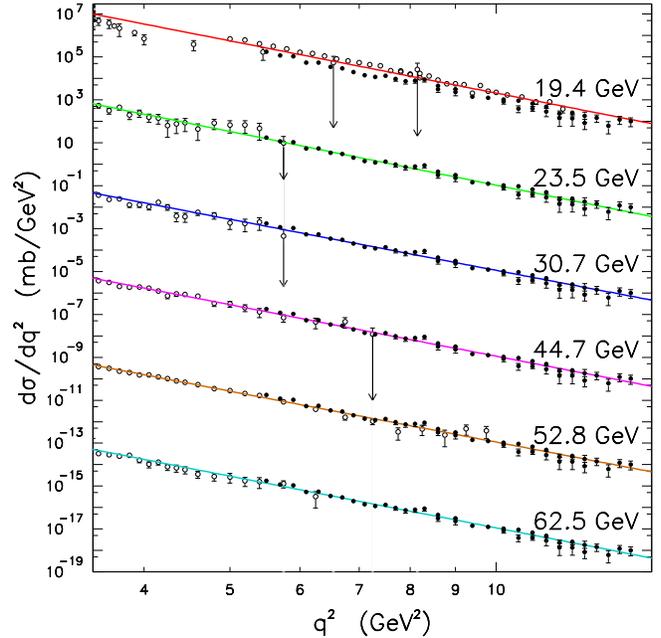}}
\caption{Addition of the data at 27.4 GeV and fit through
parametrization (\ref{eq:10}) in logarithmic scales, with cutoff at $q_{\mathrm{min}}^2$ = 3.5 GeV$^2$.
Curves and data were multiplied by factors of $10^{\pm 4}$.}
\label{fig:4}       
\end{figure}

The point is to add the data at 27.4 GeV to each set at nearby
energies, from 19.4 to 62.5 GeV and perform the fits
to each ensemble with the above parametrization. We have introduced three
cutoffs for the momentum transfer,  $q_{\mathrm{min}}^{2}$ = 3.5, 4.5 
and 5.5 GeV$^2$ and have considered either $\lambda$ = 8 (as predicted
by Donnachie and Landshoff \cite{land}) or $\lambda$ as a free fit parameter.
The numerical results of these tests, obtained through the CERN-Minuit code
\cite{minuit}, are displayed in Table~\ref{tab:3}. Figure~\ref{fig:4}
illustrates the fits in the case of the lowest cutoff, 
$q_{\mathrm{min}}^{2}$ = 3.5 GeV$^2$. 

\begin{table*}
\begin{center}
\caption{Tests on data at large momentum transfers
through parametrization (\ref{eq:10}).}
\label{tab:3}
\begin{tabular}{ccccccc}
\hline
$q_{\mathrm{min}}^2$ (GeV$^2$) & $\sqrt{s}$ (GeV) &  $DOF$ &  $\chi^2/DOF$ &  
$K$  (mbGeV$^{-2}$)        & $\lambda$ & average $\chi^2/DOF$ at ISR energies \\
\hline
    &19.4 & 82  & 23.4 & $0.2140  \pm  0.0015$ & 8 & \\
    &23.5 & 54  & 1.49 & $0.1183  \pm  0.0030$ & 8 & \\
    &30.7 & 54  & 2.24 & $0.1094 \pm  0.0027$ & 8 &  \\
    &44.7 & 57  & 1.69 & $0.1123  \pm  0.0026$ & 8 & $1.83 \pm 0.30$ \\
    &52.8 & 62  & 2.05 & $0.1040 \pm  0.0017$ & 8 & \\
    &62.5 & 55  & 1.69 & $0.1130  \pm  0.0028$ & 8 & \\
3.5 &     &     &      &                     &    &                \\
    &19.4 & 81  & 23.6 & $0.255  \pm  0.016$ & $8.087 \pm  0.030$ & \\
    &23.5 & 53  & 1.26 & $0.219  \pm  0.036$ & $ 8.313 \pm  0.086$ & \\
    &30.7 & 53  & 2.23 & $0.091  \pm  0.011$ & $7.900  \pm  0.064$ & \\
    &44.7 & 56  & 1.72 & $0.108  \pm  0.012$ & $7.978 \pm  0.060$ & $1.78 \pm 0.36$ \\
    &52.8 & 61  & 1.98 & $0.0862  \pm  0.0070$ & $7.883  \pm  0.048$ & \\
    &62.5 & 54  & 1.72 & $0.116  \pm  0.015$ & $8.012  \pm  0.067$ & \\
\hline
    &19.4 & 76  & 23.4 & $0.2172  \pm  0.0018$ & 8 & \\
    &23.5 & 44  & 1.73 & $0.1184  \pm  0.0031$ & 8 & \\
    &30.7 & 44  & 1.71 & $0.1177 \pm  0.0031$ & 8  & \\
    &44.7 & 47  & 1.61 & $0.1178  \pm  0.0030$ & 8 & $1.74 \pm 0.11$ \\
    &52.8 & 52  & 1.92 & $0.1117 \pm  0.0025$ & 8 & \\
    &62.5 & 45  & 1.74 & $0.1170  \pm  0.0030$ & 8 & \\
4.5 &     &     &      &                     &       &             \\
    &19.4 & 75  & 22.8 & $0.390  \pm  0.027$ & $8.291 \pm  0.034$ & \\
    &23.5 & 43  & 1.33 & $0.289  \pm  0.062$ & $ 8.45 \pm  0.11$ & \\
    &30.7 & 43  & 1.40 & $0.251  \pm  0.050$ & $8.38  \pm  0.10$ & \\
    &44.7 & 46  & 1.32 & $0.248  \pm  0.048$ & $8.38  \pm  0.10$ & $1.50 \pm 0.24$ \\
    &52.8 & 51  & 1.90 & $0.146  \pm  0.022$ & $8.142  \pm  0.079$ & \\
    &62.5 & 44  & 1.53 & $0.221  \pm  0.043$ & $8.320  \pm  0.099$ & \\
\hline
    &19.4 & 71  & 22.6 & $0.2102  \pm  0.0018$ & 8 & \\
    &23.5 & 39  & 1.84 & $0.1183  \pm  0.0031$ & 8 & \\
    &30.7 & 39  & 1.88 & $0.1181 \pm  0.0020$ & 8  & \\
    &44.7 & 42  & 1.76 & $0.1180  \pm  0.0031$ & 8 & $1.75 \pm 0.33$ \\
    &52.8 & 47  & 2.06 & $0.1134 \pm  0.0033$ & 8 & \\
    &62.5 & 40  & 1.82 & $0.1182  \pm  0.0031$ & 8 & \\
5.5 &     &     &      &                     &       &             \\
    &19.4 & 70  & 22.2 & $0.257  \pm  0.022$ & $8.097 \pm  0.040$ & \\
    &23.5 & 38  & 1.40 & $0.288  \pm  0.061$ & $ 8.45 \pm  0.11$ & \\
    &30.7 & 38  & 1.46 & $0.283  \pm  0.060$ & $8.44  \pm  0.11$ & \\
    &44.7 & 41  & 1.38 & $0.279  \pm  0.060$ & $8.43  \pm  0.11$ & $1.58 \pm 0.22$ \\
    &52.8 & 46  & 1.83 & $0.234  \pm  0.050$ & $8.37  \pm  0.11$ & \\
    &62.5 & 39  & 1.40 & $0.288  \pm  0.061$ & $8.44  \pm  0.11$ & \\
\hline
\end{tabular}
\end{center}
\end{table*}

These tests indicate the following features concerning the
energy dependence of each set:

(1) The data at 19.4 GeV are not compatible with the power law
and with data at 27.4 GeV since in all the cases (3 cutoffs)
$\chi^2/DOF \sim$ 20 for $\sim$ 50 $DOF$;

(2) As expected the best statistical results were obtained
with $\lambda$ as a free parameter. In this case, their values
deviate from  8 as the cutoff increases, reaching $\lambda \sim $ 8.4
for $q_{\mathrm{min}}^{2}$ = 5.5 GeV$^2$
(compatible with the numerical value presented by Faissler et al.).

(3) Each set at the ISR energy region is compatible with the
power law and with data at 27.4 GeV (the last column shows the average
$\chi^2/DOF$ at the ISR region). Although the data at 23.5 GeV cover the
region up to 5.75 GeV$^2$ and those at 27.4 GeV starts at 5.5 GeV$^2$
the fits indicate a global compatibility for cutoffs at 3.5 and 4.5
GeV$^2$. For that reason we may consider the data at 23.5 GeV as
a limit point for the beginning of the energy independence.

\begin{figure}
\resizebox{0.48\textwidth}{!}{\includegraphics{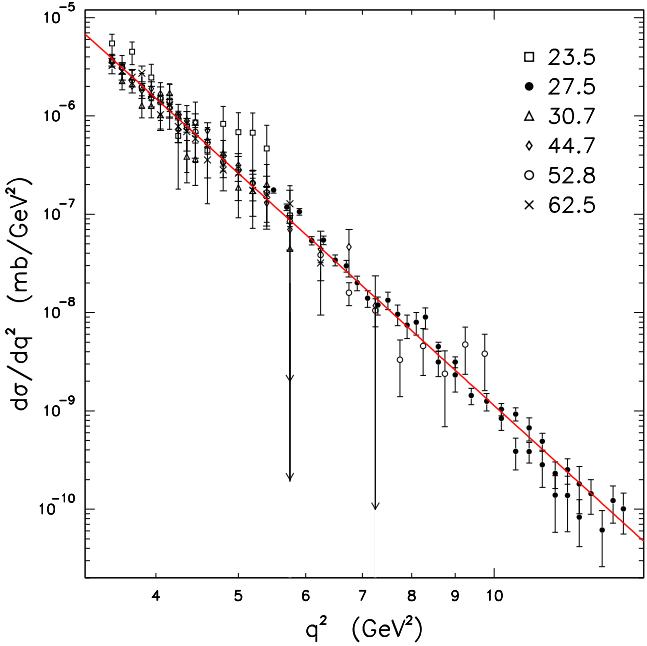}}
\caption{Fit to $pp$ differential cross section data 
from ISR together with data at 27.4 GeV (ISR + 27.4) and $q^2 >$ 3.5 GeV$^2$.}
\label{fig:5}       
\end{figure}

These conclusions can be corroborated by performing the same test
with all the ISR data together and then by adding to this ensemble
the data at 27.4 GeV. For completeness we also consider the fit
to data at 19.4 GeV alone. The results with cutoff at 3.5 GeV$^2$
are displayed in Table~\ref{tab:4}, where the above ensembles are denoted by
ISR, ISR + 27.4 and 19.4, respectively. Figure~\ref{fig:5} shows the fit result
in the case of the ensemble ISR + 27.4 and $\lambda$ as a free fit parameter.

\begin{table*}
\begin{center}
\caption{Fits through parametrization (\ref{eq:10}) to: (1) all the ISR data
(ISR); (2) all the ISR data together with data at 27.4 GeV (ISR + 27.4);
(3) data at 19.4 GeV (19.4).}
\label{tab:4}
\begin{tabular}{ccccc}
\hline
 Ensemble  &  $DOF$ & $\chi^2/DOF$ & $K$ (mbGeV$^{-2}$) & $\lambda$ \\
\hline
 ISR & $90$ & $0.97$ & $0.09635 \pm 0.00096$ & 8 \\
 & $89$ & $ 0.87$ & $0.085 \pm 0.013$ & $7.91 \pm 0.11$ \\
 ISR + 27.4  & $129$ & $1.46$ & $0.1012 \pm 0.0014$ & 8 \\
 &  $128$ & $1.38$ & $0.0798 \pm 0.0055$ & $ 7.847 \pm 0.042$\\
 19.4 & $43$ & $10.0$ & $0.2571 \pm 0.0021$ & 8 \\
 & $42$ & $11.3$ & $0.258 \pm 0.017$ & $8.003 \pm 0.032$ \\
\hline
\end{tabular}
\end{center}
\end{table*}

The main conclusion, for what follows, is that the data at 27.4 GeV
can be added to each of the 5 ISR data sets, leading to ensembles
with improved experimental information in the region of the
large momentum transfers ($q_{\mathrm{max}}^2$ = 14.2 GeV$^2$ in all the
cases), reducing
the uncertainties in the extrapolated fits. That, however, is not the
case of data at 19.4 GeV, for which $q_{\mathrm{max}}^2$ = 11.9 GeV$^2$.

\section{Improvements in the previous analysis and fit results}
\label{sec:4}

Now we first discuss some improvements introduced in the previous analyses
\cite{cmm,cm},
which are based on three aspects:
the data ensemble (selected data) and fit procedure,
the structure
of the parametrization and
the confidence intervals for the uncertainties in the fit parameters.
After that we present our new fit results.

\subsection{Data ensembles and energy independence}
\label{sec:4.1}

Based on all the information and comments presented in Sect.~\ref{sec:3},
we call the attention here to two errors appearing in \cite{cmm,cm}
and discuss the corrections needed. These concern the selected data
at 19.4 and 27.4 GeV, as well as the criterion for the energy
independence at large momentum transfer. 

First, in References \cite{cmm} and \cite{cm} the data set at 27.4
GeV was extracted from the LB tables and as we have seen, these
30 data points do not correspond to the final result with 39 points
published by Faissler et al. (Sect. \ref{sec:3.3.1}). Here we make use 
of this
latter data set.

Secondly, in \cite{cmm} the data at 19.4 GeV covers only the region up to
8.15 GeV$^2$ since the data by Faissler et al. at this energy (up to 11.9 GeV$^2$)
were not included in the analysis. Here, as referred to in Sect. \ref{sec:3.2.1}
we include all the data available at this energy.

Finally, in the fit procedure developed in \cite{cmm} the data at 27.4 GeV were added
to the data set at 19.4 GeV. However, as we have discussed, there is no statistical
justification for this addition due to energy dependence present in this region.

As we shall show in the next sections, these corrections play an important role in the 
fit results, specially in the statistical evidence for eikonal
zeros and the dependence of the position of the zeros on the energy.

\subsection{Parametrization}

In \cite{cmm} and \cite{cm} the parametrization for the real and
imaginary parts of the scattering amplitude was expressed in terms of a sum
of exponential in $q^2$ and the experimental $\rho$ value
at each energy as input. Here we use the same basic form but include also the
the total cross section as input parameter.  Specifically, 
the scattering amplitude, $F(s,q)=\Re F(s,q)+\mathrm{i}\Im F(s,q)$,
is parametrized by

\begin{eqnarray}
F(s,q) = \mu(s) \sum_{j=1}^{m} \alpha_{j} e^{-\beta_{j} q^{2}}
+
\mathrm{i}
\sum_{j=1}^{n} \alpha_{j} e^{-\beta_{j} q^{2}},
\label{eq:11}
\end{eqnarray}
where here,

\begin{equation}
\mu (s)= \frac{\rho(s)\sigma_{\mathrm{tot}}(s)}{4\pi \sum_{j=1}^{m} \alpha_{j}},
\label{eq:12}
\end{equation}
and $\rho(s)$ and $\sigma_{\mathrm{tot}}(s)$ are the experimental values
at each energy. 
In this way the
parametrization now reproduces both Eqs. (\ref{eq:5}) and (\ref{eq:6}).

\subsection{Confidence intervals for uncertainties}

Another improvement concerns the confidence level for
estimating the errors in the fit parameters (variances
and covariances). In \cite{cmm,cm} the errors correspond to an increase 
of the $\chi^2$ by one unity, which is controlled in the CERN-Minuit code
by the \textit{up} parameter, being set equal to 1. 
Depending on the number of free parameters, this fixed value implies in different
confidence level intervals, which determine the interval of the uncertainty
in each free parameter \cite{minuit}. With this procedure, any error propagation
is different for fits with different number of parameters.
Here, on the other hand, we fixed the confidence interval using
the corresponding \textit{up} value for each number of parameters.
Specifically, the errors in the
fit parameters correspond to the projection of the $\chi^2$ 
hypersurface containing 70\% of probability in each energy analyzed.

\subsection{Fit results}

Summarizing, we analyze six ensembles of data on $pp$ differential
cross sections: the set at 19.4 GeV and the five  sets at
the ISR energies with the data at 27.4 GeV added to each set.
The data cover the region above the Coulomb-nuclear interference 
and include the optical points (Table~\ref{tab:1}). The errors are the
statistical only.

Each set was fitted through parametrization (\ref{eq:11}-\ref{eq:12}), with the
experimental values of $\sigma_{\mathrm{tot}}(s)$ and $\rho(s)$  at each energy 
(Table~\ref{tab:1}), by means of the CERN-Minuit code. The best fits were obtained with
2 exponential in the real part and 4, 5 or 6 in the imaginary part
depending on the data set analyzed: $m = 2$ and $n = 4$, $5$ or $6$ in
Eq. (\ref{eq:11}). We note that the exponential terms with $j = 1$ and $j = 2$
appears in both the real and imaginary parts.

\begin{figure}
\resizebox{0.48\textwidth}{!}{\includegraphics{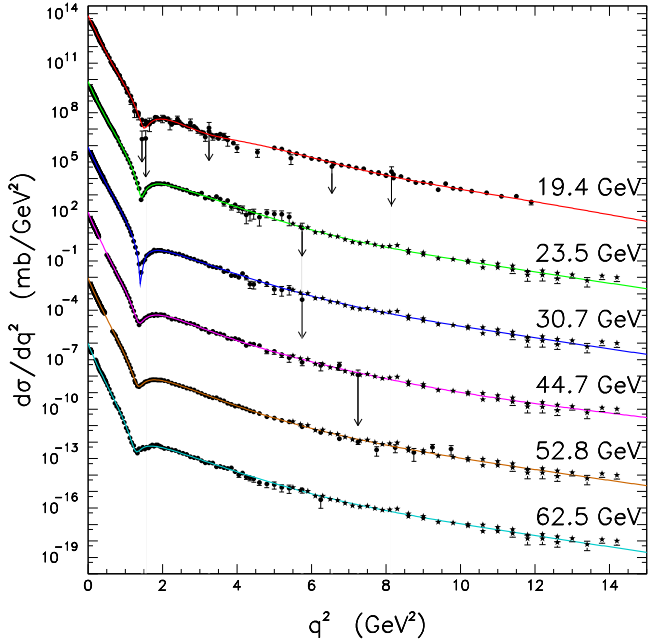}}
\caption{Results of the fits to $pp$ differential cross section
data. Curves and data were multiplied by factors of $10^{\pm 4}$.}
\label{fig:6}
\end{figure}

\begin{figure}
\resizebox{0.48\textwidth}{!}{\includegraphics{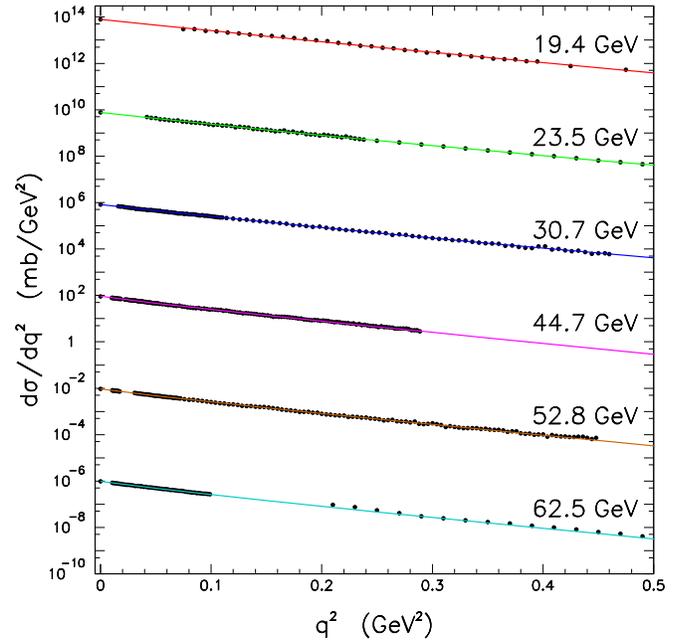}}
\caption{Results of the fit at the diffraction peak.
Curves and data were multiplied by factors of $10^{\pm 4}$.}
\label{fig:7}
\end{figure}

\begin{figure}
\resizebox{0.48\textwidth}{!}{\includegraphics{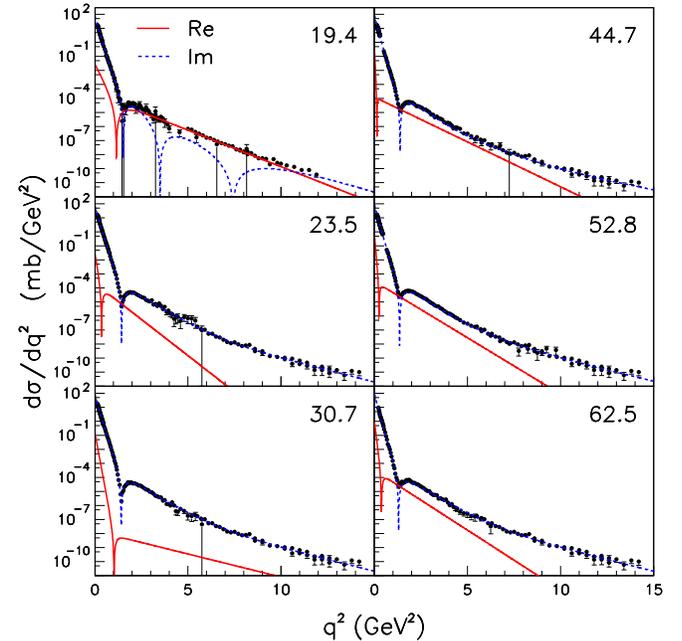}}
\caption{Contributions to the differential cross sections from
the real (dotted) and imaginary (dashed) parts of the amplitude.}
\label{fig:8}
\end{figure}

The numerical results of the fits are displayed in Table~\ref{tab:5}
together with the statistical information, including the value of the
$up$ parameter and the values of the $\chi^2/DOF$ obtained
in the previous analysis \cite{cmm}. Figures \ref{fig:6} 
and \ref{fig:7} show the fit results together
with the experimental data in the whole $q^2$ region and at the
diffraction peak, respectively. In Fig. \ref{fig:8} we display the contributions
to the differential cross section from the real and imaginary
parts of the amplitude.

\begin{table*}
\begin{center}
\caption{Fit results and statistical information from each data set: 
values of the free parameters, 
 maximum value of the momentum transfer in GeV$^2$ ($q_{\mathrm{max}}^2$),
values of the $up$ parameter for each fit (see text),
number of degrees of freedom ($DOF$) and
chi square per degree of freedom ($\chi^2/DOF$) obtained in this analysis
and that obtained in \cite{cmm}.}
\label{tab:5}
\begin{tabular}{ccccccc}
\hline
$\sqrt{s}$ (GeV):  & 19.4         & 23.5         & 30.7                   & 44.7                   & 52.8 & 62.5 \\
\hline
$\alpha_1$   & 0.1364       & -0.260       & $-1.20\times10^{-3}$   & -0.0119                & -0.0281                &  -0.042   \\
             & $\pm$ 0.0041 & $\pm$ 0.074  & $\pm0.87\times10^{-3}$ & $\pm$ 0.0024           & $\pm$ 0.0045           & $\pm$ 0.014 \\
$\alpha_2$   & -1.655       & 3.4          & 3.70                   & 0.631                  & 1.26                   & 2.20      \\
             & $\pm$ 0.066  & $\pm$ 1.3    & $\pm$ 0.49             & $\pm$ 0.090            & $\pm$ 0.13             & $\pm$ 0.61 \\
$\alpha_3$   & 3.686        & 0.25         & -0.0441                & 3.710                  & 3.631                  & 0.20      \\
             & $\pm$ 0.069  & $\pm$ 0.13   & $\pm$ 0.0063           & $\pm$ 0.053            & $\pm$ 0.060            & $\pm$ 0.28 \\
$\alpha_4$   & -1.495       & -            & -                      & -3.096                 & -3.116                 & -                    \\
             & $\pm$ 0.042  &              &                        & $\pm$ 0.050            & $\pm$ 0.056            & \\
$\alpha_5$   & 7.396        & -0.0014      & 4.51                   & 7.425                  & 6.996                  &  6.46    \\
             & $\pm$ 0.086  & $\pm$ 0.0017 & $\pm$ 0.51             & $\pm$ 0.075            & $\pm$ 0.012            & $\pm$ 0.64 \\
$\alpha_6$   & -0.1093      & 4.6          & -                      & $-0.39\times10^{-3}$   & $-1.06\times10^{-3}$   & -0.0013   \\
             & $\pm$ 0.0040 & $\pm$ 1.3    &                        & $\pm0.38\times10^{-3}$ & $\pm0.54\times10^{-3}$ & $\pm$ 0.0013 \\
$\beta_1$    & 0.6002       & 1.19         & 0.378                  & 0.736                  & 0.926                  & 0.98      \\
             & $\pm$ 0.0060 & $\pm$ 0.29   & $\pm$ 0.067            & $\pm$ 0.049            & $\pm$ 0.051            & $\pm$ 0.14 \\
$\beta_2$    & 2.762        &  8.4         & 8.18                   & 31.6                   & 16.5                   & 11.6      \\
             & $\pm$ 0.063  & $\pm$ 1.7    & $\pm$ 0.62             & $\pm$ 6.4              & $\pm$ 1.6              & $\pm$ 1.9 \\
$\beta_3$    & 2.272        & 1.31         & 0.984                  & 2.183                  & 2.217                  & 2.89      \\
             & $\pm$ 0.017  & $\pm$ 0.54   & $\pm$ 0.079            & $\pm$ 0.014            & $\pm$ 0.015            & $\pm$ 0.94 \\
$\beta_4$    & 1.770        & -            & -                      & 2.063                  & 2.126                  & -                 \\
             & $\pm$ 0.017  &              &                        & $\pm$ 0.013            & $\pm$ 0.014            & \\
$\beta_5$    & 5.864        & 0.39         & 4.21                   & 6.092                  & 5.646                  & 5.18      \\
             & $\pm$ 0.077  & $\pm$ 0.11   & $\pm$ 0.12             & $\pm$ 0.086            & $\pm$ 0.086            & $\pm$ 0.25 \\
$\beta_6$    & 0.5706       & 4.24         & -                      & 0.292                  & 0.368                  & 0.382     \\
             & $\pm$ 0.0046 & $\pm$ 0.44   &                        & $\pm$ 0.081            & $\pm$ 0.048            & $\pm$ 0.092 \\
$q_{\mathrm{max}}^2$ (GeV$^2$) & 11.9 & 14.2 & 14.2   & 14.2      & 14.2     & 14.2 \\
$up$ & 14.02    & 11.78 & 9.52     & 14.02   & 14.02  & 11.78 \\
$DOF$    & 145   & 163  & 204   & 235    & 233       & 154   \\
$\chi^2/DOF$ & 2.76     & 1.20   & 1.24    & 2.05   & 1.71   & 1.22     \\
$\chi^2/DOF$ in \cite{cmm} & 2.80  & 1.20   & 1.28    & 2.13   & 2.07   & 1.51  \\
\hline
\end{tabular}
\end{center}
\end{table*}

From Table~\ref{tab:5} we see that the values of the $\chi^2/DOF$ here obtained 
are slightly small than those presented in \cite{cmm}, except at 23.5 GeV
for which the value is the same. This slight improvement in the
statistical result may be due to the inclusion of the $\sigma_{\mathrm{tot}}$
experimental data in the parametrization at each energy.
From Fig. 8 we see that in all the cases the real part of the
amplitude presents  one zero at small values of the momentum transfers,
a result in agreement with the theorem by Martin for the real
amplitude \cite{martin}. The imaginary part develops one zero at the
ISR energies and multiple zeros at 19.4 GeV and that may be due to the fact that
these data are not normalized as in the analysis by Amaldi and Schubert. 
This effect at 19.4 did not
appear in the previous analysis \cite{cmm} due to the 
unjustified addition of data
at 27.4 GeV.

\section{Eikonal in the momentum transfer space}
\label{sec:5}

The point in the extraction of the eikonal is not only its determination
in the $q$ space, through the steps described in Sect.~\ref{sec:2}, but mainly
the estimation of the uncertainty regions by means of propagation of the errors
in the fit parameters (variances and covariances) and also the errors from 
$\sigma_{\mathrm{tot}}(s)$ and
$\rho(s)$. The problem here is that, with parametrizations like
(\ref{eq:11} - \ref{eq:12}) for the scattering amplitude (sum of exponential in $q^2$), the translation
of the eikonal from $b$-space to $q$-space, Eq.~(\ref{eq:3}), can not be analytically
performed and therefore, the standard error propagation neither. To solve
this problem 
a \textit{semi-analytical
method} was developed, which is explained in detail in \cite{cmm,cm} 
and will also be applied
in this analysis.

In what follows we shall
treat only the \textit{imaginary part of the eikonal} since,
according to our definition, Eqs. (\ref{eq:1}) and (\ref{eq:2}), it corresponds
to a real opacity function in the optical analogy. With the usual notation
we represent

\begin{equation}
\Im \chi(s,b) \equiv \Omega(s,b).
\label{eq:13}
\end{equation}
We shall also use the bracket $<\ >$ to denote two dimensional Fourier
transform with azimuthal symmetry, so that the translations between
$q$ and $b$ spaces will be expressed by

\begin{eqnarray}
\Omega(s,b) =  <\tilde\Omega(s,q)> =
\int_{0}^{\infty} q \mathrm{d}qJ_{0}(qb) \tilde\Omega(s,q),
\label{eq:14}
\end{eqnarray}

\begin{eqnarray}
\tilde\Omega(s,q) =  <\Omega(s,b)> =
\int_{0}^{\infty} b \mathrm{d}bJ_{0}(qb) \Omega(s,b).
\label{eq:15}
\end{eqnarray}

\subsection{Semi-analytical method}

As shown in \cite{cmm}, taking into account the error propagation from 
the fit parameters
it is possible to approximate the imaginary part of the
eikonal in Eq.~(\ref{eq:8}) by

\begin{eqnarray}
\Omega(s,b)
\approx
\ln\left[\frac{1}{1-\mathrm{Re}\ \Gamma(s,b)}\right] 
\label{eq:16}
\end{eqnarray}
and the same is valid in the present analysis. Expanding this equation we obtain

\begin{equation}
\Omega(s,b)=\mathrm{Re}\ \Gamma(s,b)+R(s,b),
\label{eq:17}
\end{equation}
where $R(s,b)$ represents the remainder of the series:

\begin{equation}
R(s,b)=\ln\left[\frac{1}{1-\mathrm{Re}\ \Gamma(s,b)}\right]-
\mathrm{Re}\ \Gamma(s,b).
\label{eq:18}
\end{equation}

Up to this point the errors of fit parameters from $\mathrm{Im}\ F$
can be propagated to $\mathrm{Re}\ \Gamma$ by Eq.~(\ref{eq:7}) and to $R(s,b)$,
Eq.~(\ref{eq:18}).
The next step concerns the translation of Eq.~(\ref{eq:17}) 
from $b$-space to $q$-space, Eq.~(\ref{eq:15}). Applying the
Fourier transform in (\ref{eq:17}) we obtain

\begin{equation}
\tilde\Omega(s,q)
= \mathrm{Im}\ F(s,q)
+ \tilde R (s,q).
\label{eq:19}
\end{equation}
As commented on before, the point here is that due to the structure of the 
parametrization,
the translation from $R(s, b)$ to $\tilde R(s, q)$ can not be
performed in an analytical way and as consequence 
nor can the error
propagation. The \textit{semi-analytical method} introduced in \cite{cm}
address this question through the following procedure. We first generate an 
ensemble of
numerical points
$R(s,b)$ through Eq.~(\ref{eq:18}),
with propagated errors $\pm\Delta R(s,b)$
 and then fit this ensemble by a 
sum of gaussians in
$b$, in practice with six terms:

\begin{equation}
R_{\textrm{fit}}\:(s,b)=\sum_{j=1}^{6} A_{j}\e^{-B_{j}b^{2}}. 
\label{eq:20}
\end{equation}

In this way, not only $\tilde R(s, q)$ can be evaluated 
through the Fourier transform of the above formula, but also
the errors from the fit parameters $A_{j}$, $B_{j}$,
can be analytically propagated providing $\Delta \tilde R$ and, through 
Eq.~(\ref{eq:19}),
$\Delta\tilde\Omega(s,q)$. As discussed in \cite{cmm}, this method allows the 
study of several aspects of the eikonal in the momentum transfer space.
In this work we shall focus only on the investigation of eikonal zeros (change of sign).

\subsection{Eikonal zeros}

\begin{figure}
\resizebox{0.48\textwidth}{!}{\includegraphics{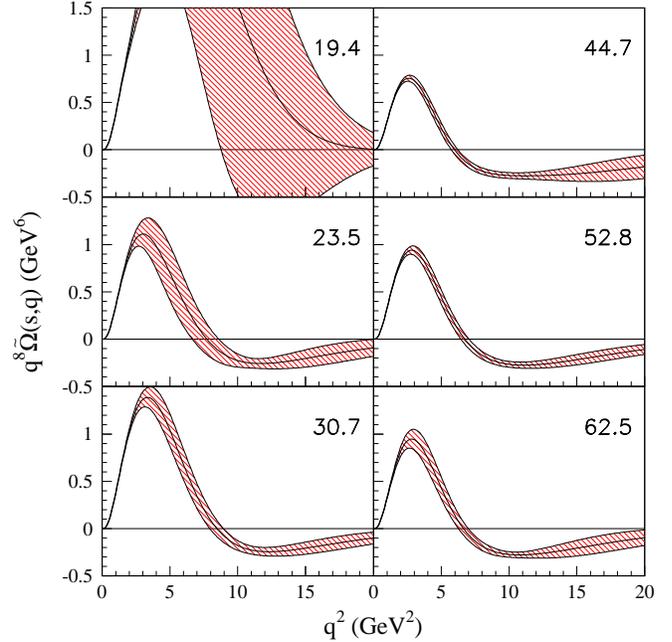}}
\caption{Imaginary part of the Eikonal in the momentum transfer space multiplied by
$q^8$ and uncertainty regions from error propagation.}
\label{fig:9}
\end{figure}

A review on previous indication of eikonal zeros, with complete
references to outstanding results can be found in \cite{cmm}.
Here we use the semi-analytical method in order to investigate
the eikonal zeros and the associated uncertainties. 
As in \cite{cmm,cm} that can be done through plots of
$q^8$ times $\tilde\Omega(s,q) \pm \Delta \tilde\Omega(s,q)$ as function of the
momentum transfer as shown in Fig.~\ref{fig:9}.
We consider as statistical evidence of a change
of signal only the cases in which the uncertainty region above the
central value is below the zero.
With this criterion, from Fig. \ref{fig:9},  we have evidence for the change of
sign at all the ISR energies, but, different from the result
obtained in \cite{cmm}, not  at $\sqrt s$ = 19.4 GeV. 

\begin{figure}
\resizebox{0.48\textwidth}{!}{\includegraphics{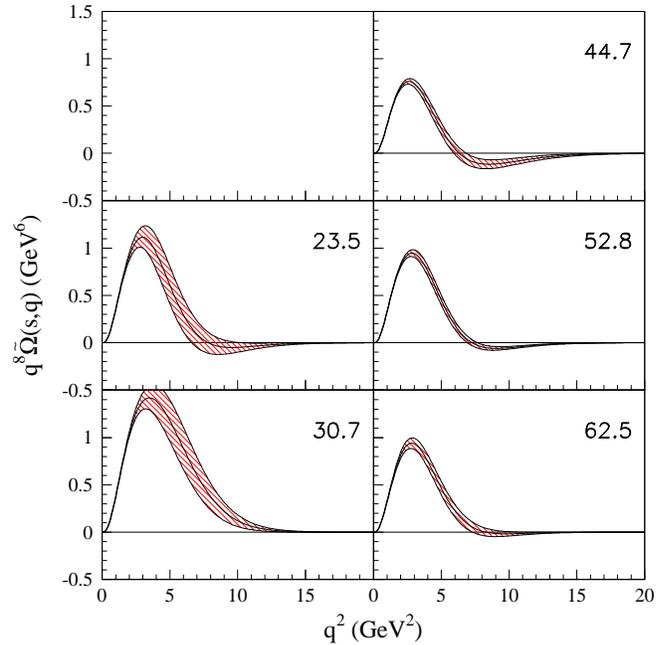}}
\caption{Same as Fig.~\ref{fig:9} from fits to the ISR data sets without adding the data 
at 27.4 GeV.}
\label{fig:10}
\end{figure}

We recall that in \cite{cmm} the data at 27.4 GeV was added to those
at 19.4 GeV and that is not the case here. This suggest the importance
of data at large momentum transfer in statistical identification of a
zero in the eikonal. This aspect can also be corroborated if we consider
fits only to the original ISR data sets, that is, without adding the data at
27.4 GeV. The results with parametrization 
(\ref{eq:11}-\ref{eq:12}) are displayed in Fig.~\ref{fig:10},
from which we see that except for the data at 44.7 and  52.8 GeV 
no evidence
of zeros can be inferred and these two sets just correspond to those
with the largest interval in the momentum transfer with available
data (Fig.~\ref{fig:1}).

\begin{figure}
\resizebox{0.48\textwidth}{!}{\includegraphics{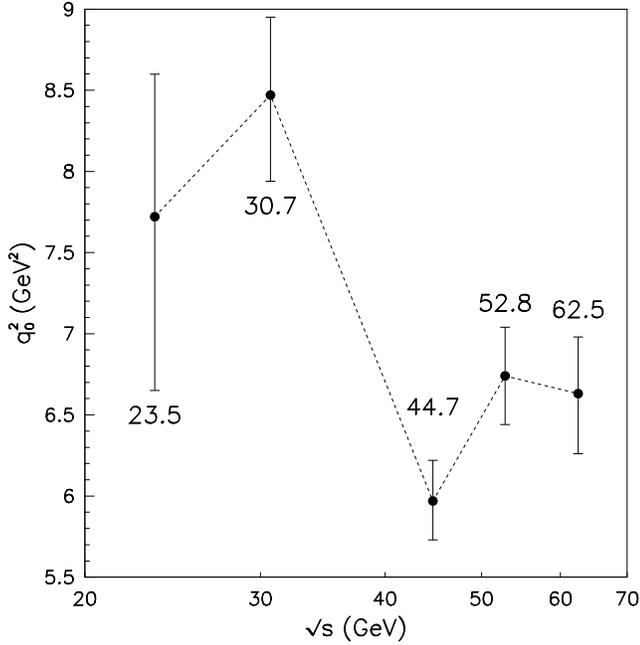}}
\caption{Position of the eikonal zero and uncertainties
in terms of the energy (Table~\ref{tab:6}).}
\label{fig:11}
\end{figure}

From the plots in Fig.~\ref{fig:9} 
and the NAG routine (C05ADF), we can determine the position of the zeros and
the uncertainties associated with each central value by means of the
extrema intervals of the propagated errors (asymmetrical). The numerical
results extracted in this way are shown in Table~\ref{tab:6}, where
$q_0^2$ indicates the central value of the zero and $+ \Delta q_0^2$
and $- \Delta q_0^2$ the asymmetrical uncertainties at 
the right and the left of the 
central value, respectively.
These numerical values are plotted in Fig.~\ref{fig:11}, where
the lines connecting the central values were drawn only to guide the eye.

\begin{table}
\begin{center}
\caption{Position of the eikonal zero ($q_0^2$) and 
the asymmetrical uncertainties ($+ \Delta q_0^2$
and $- \Delta q_0^2$) in terms of the energy.}
\label{tab:6}
\begin{tabular}{cccc}
\hline
$\sqrt{s}$ (GeV) & $q_0^2$ & $- \Delta q_0^2$ & $ + \Delta q_0^2$ (GeV$^2$) \\
\hline
23.5 & 7.72  & 1.07 & 0.88\\
30.7 & 8.54 & 0.99 & 0.80 \\
44.7 & 5.83 & 0.15 & 0.16 \\
52.8 & 6.74 & 0.64 & 0.60 \\
62.5 & 6.63 & 0.37 & 0.35 \\
\hline
\end{tabular}
\end{center}
\end{table}

Despite the statistical evidence for the change of sign of the eikonal
at the ISR energy region, these results do not allow one to extract a
quantitative correlation between the position of the zero, $q_0^2$,
and the energy. However, we can outline the following quantitative
features:

\vspace{0.2cm}

\noindent
(1) For the lower energies ($\sqrt s$ = 23.5 and 30.7 GeV$^2$)
$q_0^2 \sim $ 8 GeV$^2$ and for the higher energies (44.7, 52.8
and 62.5 GeV) $q_0^2 \sim $ 6 GeV$^2$, suggesting a decreasing
in the position of the zero as the energy increases.

\vspace{0.2cm}

\noindent
(2) Fits to the data on the zero position, Table~\ref{tab:6}, with a linear function,
$q_0^2 = a + b\ln s$ gives
$q_0^2 = (12.3 \pm 2.9) - (0.74 \pm 0.37) \ln s$, with
$\quad \chi^2/DOF =$ 5.3,
if the largest errors are used and
$q_0^2 = (12.8 \pm 2.6) - (0.81 \pm 0.3) \ln s$,
with $\quad \chi^2/DOF =$ 6.0, 
in the case of the smallest errors. Since the errors in the slopes are
about 50 \% of the central value, these results also suggest a
decreasing in $q_0^2$ as the energy increases.

\vspace{0.2cm}

\noindent
(3) If we assume $b=0$ in the above parametrizations (null slope)
we obtain, respectively,
$q_0^2 = 6.60 \pm 0.16$ GeV$^2$, $\chi^2/DOF =$ 5.0
and
$q_0^2 = 6.61 \pm 0.15$ GeV$^2$, $\chi^2/DOF$= 6.0.

\vspace{0.2cm}

\noindent
(4) The average of only the central values gives
$\bar{q_0^2} = 7.04 \pm 1.08$ GeV$^2$.

\vspace{0.2cm}

From these numerical results we can only infer the evidence of the change
of sign in the eikonal in the region 6 - 8 GeV$^2$, what can roughly
be represented by the value:

\begin{equation}
\bar{q_0^2} = 7.0 \pm 1.0\ \mathrm{GeV}^2. 
\label{eq:21}
\end{equation}

We note that the conclusion that the position of the zero decreases
with the increasing of the energy, inferred in \cite{cmm}, was based on the
position of the zero at $\sqrt s$ = 19.4 GeV, namely
$q_0^2 \approx$ 9 GeV$^2$
(see Fig.~\ref{fig:15} in that reference). However, without adding the data
at 27.4 GeV, as we did here, we can not infer this result.
In what
follows we discuss the implication of this zero in the
phenomenological context.

\section{Phenomenological implication of the eikonal zeros}
\label{sec:6}

First, an important observation. Despite the detailed model independent analysis
here developed, it should be stressed that we do not present
$the$ empirical result, but $an$ empirical result. In fact, even with the 
justified strategy of adding data at large momentum transfer, the fit procedure
has, in principle, an infinity number of solutions. In our case, this drawback is mainly
associated with the lack of knowledge of the contributions from real
and imaginary parts of the amplitude beyond the forward direction,
which represents a serious challenge in any inverse scattering problem.
For that reason, in what follows, we shall base our general discussion
in qualitative aspects, treating also some quantitative features
but without going into details.

Summarizing, our model independent result for the imaginary part of the
eikonal in the $q$ space indicates that, at the ISR region, the eikonal
is positive up to $q_0^2 \sim$ 7 GeV$^2$, changes sign at this point, has 
a negative minimum above the zero position and then goes to zero
through negative values (Fig.~\ref{fig:9}). As already discussed by
Kawasaki, Maheara and Yonegawa \cite{kmy} this behavior suggests
two distinct dynamical contributions in the diffractive regime:
an interaction with long range (positive eikonal below the
zero) and another with short range (negative eikonal above the
zero).
In this Section we discuss some implication of this behavior. 
We first treat empirical results related with the
eikonal and the scattering amplitude (Sect.~\ref{sec:6.1}) and then the
implication on the zero in terms of eikonal models 
(Sect.~\ref{sec:6.2}) and form factors (Sect.~\ref{sec:6.3}).

\subsection{Eikonal and scattering amplitude}
\label{sec:6.1}

One of the most important label of elastic hadron scattering
as a diffractive process is the diffraction pattern in the
differential cross section: the peak, the dip and the smooth
decrease at large momentum transfer. It is generally accepted
that the dip at $q^2 \sim$ 1.5 GeV$^2$ 
(Figures \ref{fig:1} and \ref{fig:2}) is due
to a change of sign (zero) in the imaginary part of the
amplitude and that the dip is filled up by the real part of
the amplitude. That is, at least, what our fit results indicate
(Fig.~\ref{fig:8}). Therefore it may be worthwhile to examine possible
connections between the zeros in the amplitudes and in the eikonal
(imaginary parts). Several interesting aspects of this subject
have already been discussed by Kawasaki, Maehara and Yonezawa
\cite{kmy}; here we focus only on our empirical results.

By expanding the exponential term in Eq.~(\ref{eq:1}) we obtain for the
imaginary part of the amplitude

\begin{eqnarray}
& &\Im F(s, q) = \mathrm{Im} <\chi(s,b)> +
\frac{1}{2!} \mathrm{Re} <\chi^2(s,b)> \nonumber \\
&-& \frac{1}{3!} \mathrm{Im} <\chi^3(s,b)> -
\frac{1}{4!} \mathrm{Re} <\chi^4(s,b)> + .... 
\label{eq:22}
\end{eqnarray}
Therefore, in principle, the zero in $\Im F(s, q)$
can be generated either by a zero in $\tilde\Omega(s,q)
= \mathrm{Im} <\chi(s,b)>$ or 
by the terms with alternating signs in the series
\cite{kmy}. Obviously the difference
between $\Im F$ and $\tilde\Omega$ (the leading term)
comes from the contribution of the reminder of the series. 

\begin{figure}
\resizebox{0.48\textwidth}{!}{\includegraphics{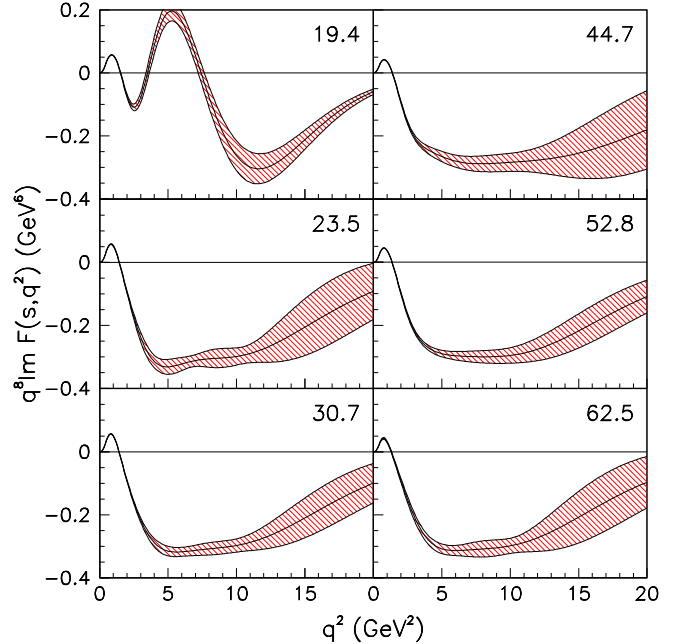}}
\caption{Fit results for the imaginary part of the amplitude, 
multiplied by $q^{8}$ and uncertainty regions from error
propagation (analogous to Fig.~\ref{fig:9} for the imaginary part
of the eikonal).}
\label{fig:12}
\end{figure}

\begin{figure}
\resizebox{0.48\textwidth}{!}{\includegraphics{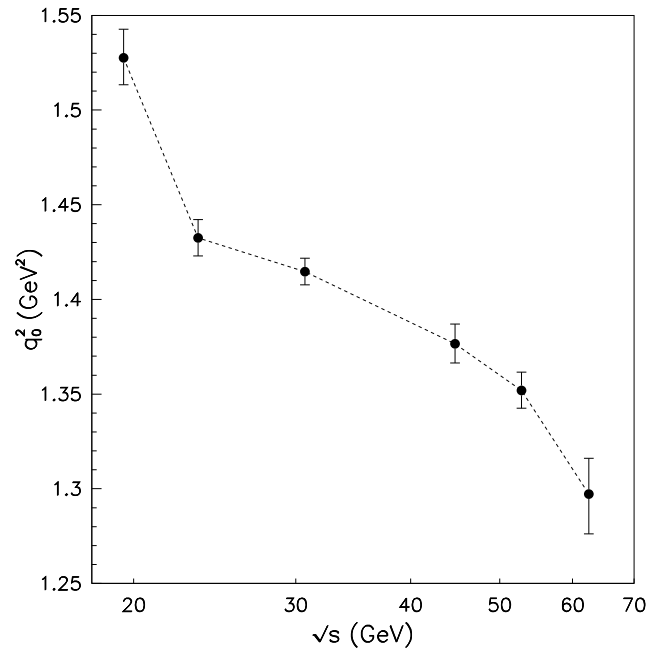}}
\caption{Position of the zero in the imaginary part of the
amplitude. At 19.4 GeV the value corresponds to the first zero
(Table~\ref{tab:7}).}
\label{fig:13}
\end{figure}

Quantitative information on this respect can be obtained directly from our fit results
and by comparing both quantities (amplitude and eikonal). To this end we consider, 
as in the case of the eikonal in the $q$-space, the product of $q^8$ by 
$\Im F(s, q)$ for all the energies analyzed as shown in Fig.~\ref{fig:12}.
With this we can
determine the position of the zero in the amplitude together with 
the propagated uncertainties. The results are displayed in Fig.~\ref{fig:13}
and Table~\ref{tab:7} (where the value at 19.4 GeV corresponds to the first zero
only).

\begin{table}
\begin{center}
\caption{Position of the zero in the imaginary part of the amplitude
(first zero in the case of 19.4 GeV).}
\label{tab:7}
\begin{tabular}{cccc}
\hline
$\sqrt{s}$ (GeV) & $q_0^2$ & $- \Delta q_0^2$ & $ + \Delta q_0^2$ (GeV$^2$) \\
\hline
19.4 & 1.528 & 0.014 & 0.015 \\
23.5 & 1.4325  & 0.0095 & 0.0097 \\
30.7 & 1.4147 & 0.0071 & 0.0071 \\
44.7 & 1.377 & 0.010 & 0.010 \\
52.8 & 1.3520 & 0.0094 & 0.0097 \\
62.5 & 1.297 & 0.021 & 0.019 \\
\hline
\end{tabular}
\end{center}
\end{table}

These results allow to extract the following empirical features:

\vspace{0.2cm}

(1) Concerning the position of the zero in the amplitude and in the eikonal,
Tables \ref{tab:6} and \ref{tab:7} and 
Figs. \ref{fig:11} and \ref{fig:13} show that there is no correlation
at all between them: in contrast with the position of the zero in the 
amplitude, which systematically decreases as the energy increases (an effect related to the
well known shrinkage of the diffraction peak), the eikonal zero
has an approximately constant position in the energy region investigated.

\vspace{0.2cm}

(2) At 19.4 GeV, from Fig.~\ref{fig:9} for the eikonal, we see that
it is not possible to identify a change of sign on statistical
grounds (uncertainties below the zero), nor in terms of the
central value neither since it goes asymptotically to zero through positive values.
From Fig.~\ref{fig:12}, the corresponding imaginary part of the amplitude
presents multiple zeros (three at finite $q^2$-values and one
asymptotically, through negative values).

\vspace{0.2cm}

(3) At the ISR energy region, from Fig.~\ref{fig:9}, we have evidence for the
change of signal (one zero) and the corresponding imaginary
part of the amplitudes, Fig.~\ref{fig:12}, also present only one change of signal at
fixed $q^2$, going to zero through negative values.

\vspace{0.2cm}

The last two features suggest that a positive-definite eikonal in
the $q$ space originates in multiple dips in the corresponding
differential cross section (zeros in the imaginary part of the
amplitude); on the other hand, an eikonal with one change of signal, gives rise
to only one dip and a smooth decrease at large momentum transfers.
In what follows we discuss these effects in the context
of some eikonal models.

\subsection{Some representative eikonal models}
\label{sec:6.2}

In order to illustrate the empirical features described above, we have
chosen some representative and popular eikonal models, characterized
by parametrizations with and without zero in the imaginary part
of the eikonal in the $q$ space. We first review some  
aspects of each model we are interested in 
(Sects. \ref{sec:6.2.1} - \ref{sec:6.2.3})
and then discuss the 
connections with the empirical results (Sect. \ref{sec:6.2.4}). 

\subsubsection{Models without eikonal zero}
\label{sec:6.2.1}

Representatives of this class are the historical Chou-Yang model 
\cite{wy,by,dl,cy,cy90} 
and some recent QCD-inspired models \cite{bghp,lmmmn}.
Due to the importance of the connections between eikonal and form factors
and for further discussion (Sect. \ref{sec:6.3}), we recall some details of the 
droplet model by Yang and collaborators and only briefly quote some 
inputs of interest in QCD models.

\vspace{0.2cm}

\noindent
$\bullet$ \textit{Chou-Yang model}

\textit{Basic concepts}. In this model, the internal structure of a hadron is assumed to 
be described
by a {\it density of opaqueness} $\rho(x,y,z)$ and in a collision,
relativistic effects imply in a contraction of the extended object, so that,
in the center-of-mass frame each hadron ``sees'' the other as a
two-dimensional matter distribution, 

\begin{eqnarray}
D(x,y) = \int_{-\infty}^{+\infty} \rho(x,y,z) \mathrm{d}z, \nonumber
\end{eqnarray}
where $x$ and $y$ lie on the impact parameter plane and $z$ is the coordinate
perpendicular to it. According to the optical analogy, the {\it resultant
opaqueness} (the imaginary part of the eikonal) in the collision of hadrons $A$
and $B$ is assumed to be the overlapping (convolution) of the matter distributions,

\begin{eqnarray}
\Omega_{AB}(s,b) &=& C_{AB}\int \mathrm{d}^2 \vec{b'} D_A(|\vec{b'}|) D_B(|\vec{b'}
- \vec{b}|) \nonumber \\ 
&\equiv& C_{AB}\ D_A \otimes D_B,
\label{eq:23}
\end{eqnarray}
where $C_{AB}$ is impact parameter independent (it depends on the energy), and
$D_{A,B}$ are connected to the \textit{hadronic matter form factors},

\begin{eqnarray}
G_{A,B}(\vec{q}) = \int \e^{ \mathrm{i}\vec{q} \cdot \vec{r} }
\rho_{A,B}(\vec{r}) \mathrm{d}\vec{r}, 
\label{eq:24}
\end{eqnarray}
through the Fourier transform,

\begin{equation}
D_{A,B}(b) = < G_{A,B}(q) >.
\label{eq:25}
\end{equation}
From the convolution theorem, we obtain the formal expression of the eikonal
in the impact parameter space:

\begin{equation}
\Omega_{AB}(s,b) = C_{AB}(s) < G_A(q) G_B(q)>.  
\label{eq:26}
\end{equation}

\textit{The Wu-Yang conjecture}.
In 1965, based on heuristic arguments, Wu and Yang speculated that
the elastic $pp$ differential cross sections might be proportional to the
fourth power of the \textit{proton charge form factor} \cite{wy}, that is
the form factor measured in electron-proton scattering. The connection
with this power of the form factor is obtained, in the above model,
by considering the first order expansion
of the eikonal, Eqs. (\ref{eq:22}) and (\ref{eq:26}), since for the proton case: 

\begin{eqnarray}
\Im F(s, q) \propto G_p^2.
\label{eq:27}
\end{eqnarray}

Several electromagnetic form factors and inverse
scattering problems were discussed in the subsequent years \cite{by,dl,cy},
including the traditional dipole parametrization for the Sachas
electric form factor \cite{dl}

\begin{eqnarray}
G_D(q) = \frac{1}{[1 + q^2/\mu^2]^2}, \qquad \mu^2 = 0.71\ {\rm GeV}^2.
\label{eq:28}
\end{eqnarray}
In this case, from Eq.~(\ref{eq:26}), the opacity in the impact parameter space 
for $pp$ scattering reads

\begin{eqnarray}
\Omega(s,b) = C(s) \frac{(\mu b)^3}{8} K_3(\mu b),
\label{eq:29} 
\end{eqnarray}
where $K_3$ is a modified Bessel function. With inputs
like that, the  {\it absorption factor} $C_{AB}(s)$
is the only free parameter, determined from the experimental value of the total
cross section at each energy. The main realization of this model was the
prediction of the diffraction pattern in $pp$ differential cross section and
the correct position of the dip, as experimentally observed later.

However, although efficient in the description of the experimental data,
the strong conjecture by Wu and Yang, correlating the \textit{hadronic matter
form factor} with the \textit{electric form factor}, can not be proved
or disproved in the phenomenological context. We shall return to this fundamental point
in Sect. \ref{sec:6.3},
when discussing recent results on the proton electric form factor.

\vspace{0.2cm}

\noindent
$\bullet$ \textit{QCD-Inspired models}

In this class of model \cite{bghp,lmmmn} 
the even eikonal is expressed as a sum of three contributions, from
gluon-gluon ($gg$), quark-gluon ($qg$) and quark-quark ($qq$) interactions,

\begin{eqnarray}
\chi^+(s,b) = \chi_{gg}(s,b) + \chi_{qg}(s,b) + \chi_{qq}(s,b), \nonumber
\end{eqnarray}
which individually factorize in $s$ and $b$,

\begin{eqnarray}
\chi_{ij}(s,b) = \mathrm{i} \sigma_{ij}(s) w(b, \mu_{ij}), \nonumber
\end{eqnarray}
where $ij$ stands for $gg$, $qg$ and $qq$.
The {\it impact parameter distribution function} for each process comes from
convolution involving dipole form factors, in the same way as in the Chou-Yang model,
Eqs. (\ref{eq:23}) and (\ref{eq:28}),
but at the elementary level:

\begin{eqnarray}
w_{ii}(b, \mu_{ii})  = \int \mathrm{d}^2\vec{b'} D_{i}(|\vec{b'}|)
\ D_{i}(|\vec{b'} - \vec{b}|), \nonumber
\end{eqnarray}

\begin{eqnarray}
G_{ii}(b, \mu_{ii}) =  \left< \frac{1}{[1 + q^2/\mu_{ii}^2]^2} \right>, \nonumber
\end{eqnarray}
so that

\begin{eqnarray}
w_{ii}(b, \mu_{ii}) = \frac{[\mu_{ii} b]^3}{8} K_3(\mu_{ii} b),
 \nonumber 
\end{eqnarray}
where, for $i \not= j$:

\begin{eqnarray}
\mu_{ij} \equiv \sqrt{\mu_{ii} \mu_{jj}}.  \nonumber
\end{eqnarray}

Therefore, in the momentum transfer space, the imaginary part of the
eikonal has the same structure of the Chou-Yang model, with the
dipole parametrization and the scale
factors $\mu_{ii}$, $i = g, q$ as free fit parameters, depending
also on the elementary process ($qq$ or $gg$).
With several other ingredients this class of model allows for good descriptions
of the forward data and differential cross section data at small
momentum transfers \cite{bghp,lmmmn}.

\subsubsection{Hybrid model}

For further discussion we also recall a particular model with 
different
parametrizations for $pp$ scattering at the ISR region and $\bar{p}p$ scattering
at the Collider energies, the former using an eikonal with multiple zeros
and the later without zero. The model, developed by Glauber and Velasco
\cite{gv1,gv2}, is based on Glauber's multiple diffraction formalism which,
in leading order, introduces the following expression for the
eikonal \cite{glauber}

\begin{eqnarray}
\tilde\chi(s, q) = \sum_{i=1}^{N_A} \sum_{j=1}^{N_B} G_A G_B f_{ij},
\nonumber
\end{eqnarray}
where $G_A$ and $G_B$ are the \textit{hadronic} form factors, 
$N_A$ and $N_B$ the number of constituents in each hadron and 
$f_{ij}$ the individual \textit{elementary} scattering amplitudes between
the constituents (parton-parton scattering amplitudes).
In the case that the elementary amplitudes
can be considered to be the same, denoted by $f$, and that $N_A N_B \equiv N$, 
we have for the imaginary part

\begin{equation}
\tilde\Omega(s, q) = N G_A G_B \mathrm{Im} f.
\label{eq:30}
\end{equation}

In the Glauber-Velasco version \cite{gv1,gv2} use is made of the Felst 
as well as the
Borkowski-Simon-Walther-Wendling (BSWW) form factors (no zeros), together with the following 
parametrization for the imaginary part of the elementary amplitude in the
case of $\bar{p}p$ scattering at 546 GeV \cite{gv1}:

\begin{eqnarray}
f(q) = \frac{1}{[1 + q^2/a^2]^{1/2}}. \nonumber
\end{eqnarray}
Therefore, the eikonal presents no zero. For $pp$ scattering at 23.5
a phase factor was introduced,

\begin{eqnarray}
f(q) = \frac{\exp{\{\mathrm{i}[b_1 q^2 + b_2 q^4]\}}}{[1 + q^2/a^2]^{1/2}}
\nonumber
\end{eqnarray}
and in this case both the real an imaginary parts of the eikonal
present multiple zeros. For further reference we recall that the data 
cover the region up to 5.5 GeV$^2$ ($pp$, 23.5 GeV) and
1.6 GeV$^2$ ($\bar{p}p$, 546 GeV), that is, not large values
of the momentum transfer.

\subsubsection{Models with eikonal zero}
\label{sec:6.2.3}

This is a restrict class of eikonal models.
We shall consider here the impact parameter picture by Bourrely, Soffer and Wu
and a geometrical or multiple diffraction approach.

\vspace{0.2cm}

\noindent
$\bullet$ \textit{Bourrely-Soffer-Wu model}

The impact parameter picture by Bourrely-Soffer-Wu (BSW) \cite{bsw,bsw03} is the most
popular and, to our knowledge, the first model to consider an eikonal zero in the momentum transfer
space. In this model the eikonal in the impact parameter space
is expressed as a sum of two terms

\begin{eqnarray}
\chi(s,b) = R(s,b) + H(s,b), \nonumber
\end{eqnarray}
where the first term is a Regge background, which takes into account the differences
between  $pp$ and $\bar{p}p$ scattering and is parametrized as

\begin{eqnarray}
<R(s,b)> = [ c_+ + c_- \e^{-\mathrm{i}\pi \alpha(q^2)} ] s^{\alpha(q^2)}, 
\nonumber
\end{eqnarray}

\begin{eqnarray}
\alpha(q^2) = \alpha_0 - \alpha'q^2, \qquad q^2 = -t. \nonumber
\end{eqnarray}
The second term, responsible for the diffractive component (pomeron exchange),
is the same for $pp$ and $\bar{p}p$ and factorizes in $s$ and $b$:

\begin{eqnarray}
H(s,b) = S(s) T(b). \nonumber
\end{eqnarray}
The energy-dependent term comes from the massive QED and is parametrized in a
crossing symmetric form

 \begin{eqnarray}
S(s) = \frac{s^c}{\ln^{c'} s} + \frac{u^c}{\ln^{c'} u}, \nonumber
\end{eqnarray}
where $u$ is the third Mandelstam variable.
Finally, the impact parameter dependence, which is our interest, is 
also inspired in the geometrical picture
through the convolution

\begin{eqnarray}
T(b) = k D_A \otimes D_B = k< G^2(q^2) >. 
\label{eq:31}
\end{eqnarray}
Here, however, the form factor is parametrized as a product of two simple
poles multiplied by a function with a zero in the momentum
transfer space,

\begin{eqnarray}
G(q^2) = \frac{1}{[1 + q^2/\alpha^2]} \frac{1}{[1 + q^2/\beta^2]} 
\sqrt{\frac{1 - q^2/q_0^2}{1 + q^2/q_0^2}},
 \label{eq:32}
\end{eqnarray}
where $k$, $\alpha^2$, $\beta^2$ and $q_0^2$ are free fit parameters.
The function on the right, with a zero at $q^2 = q_0^2$,
was introduced to account for possible
differences between the electromagnetic and hadronic form factors, as well as
to correct the dip position \cite{bsw}. In the last analysis by Bourrely, Soffer and Wu the
position of the zero was inferred to be at \cite{bsw03}

\begin{eqnarray}
q_0^2 \sim 3.45\ \mathrm{GeV}^2. 
\label{eq:33}
\end{eqnarray}

\vspace{0.2cm}

\noindent
$\bullet$ \textit{A multiple diffraction model}

Without a theoretical basis as in the case of the BSW model, a multiple
diffraction model (Glauber context), introduced
in 1988 \cite{md,bcmp}, makes use of the following parametrizations for the
eikonal in Eq.~(\ref{eq:30})

\begin{eqnarray}
\tilde\Omega(s,q) = C(s) G^2(s,q) \mathrm{Im}\ f(q), 
\label{eq:34}
\end{eqnarray}
with

\begin{eqnarray}
G(s, q) &=& \frac{1}{[1 + q^2/\alpha^2(s)]} \frac{1}{[1 + q^2/\beta^2]},
\label{eq:35}
\end{eqnarray}

\begin{eqnarray}
\mathrm{Im} f(s,q) &=&  \frac{1 - q^2/q_0^2}{1 + [q^2/q_0^2]^2},
\label{eq:36}
\end{eqnarray}
where the $N$ factor has been included in $C(s)$.
The mathematical structure is very similar to the geometrical ansatz introduced by BSW,
except for the dependence of $\alpha^2$ on the energy and the square in the
$q^2/q_0^2$ term in the denominator. The reason for this square is explained
and discussed in \cite{mprd}. By means of suitable phenomenological parametrizations
for $C(s)$ and $\alpha^2(s)$ and for 

\begin{eqnarray}
q_0^2 = 8.20\ \mathrm{GeV}^2, 
 \label{eq:37}
\end{eqnarray}
good descriptions of the
experimental data on elastic $pp$ and $\bar{p}p$ scattering, above 10 GeV,
have been obtained ($\beta_{pp}^2 = 1.80$ GeV$^2$ and  $\beta_{\bar{p}p}^2 = 1.55$ GeV$^2$);
the real part of the amplitude can be evaluated either through the
Martin formula \cite{md,mblois,mcjp} or by means of derivative dispersion
relations applied at the elementary level \cite{mm}.

In the geometrical context the $\alpha^2$ dependence means hadronic form
factors depending on the energy, a hypothesis or procedure that was also used in 1990 by Chou and Yang
\cite{cy90}. A theoretically improved version of this 
multiple diffraction model, including dual and pomeron
aspects, is presented in \cite{covoetal}.

\subsubsection{Discussion}
\label{sec:6.2.4}

\noindent
$\bullet$ \textit{General aspects}

As is known from the original papers, all the above models
\textit{without zero} in the eikonal can only describe the differential
cross section data at small values of the momentum transfer,
typically below $q^2 \sim $ 2 GeV$^2$ (for example, QCD inspired models
\cite{bghp,lmmmn}). 
Above this region, theoretical curves present multiple dips which
are not present in the experimental data (for example, Chou-Yang model
\cite{cy,cy90}
and Glauber-Velasco model at Collider energies \cite{gv1}).

On the other hand, models with one zero in the eikonal are able to describe
quite well the differential cross section data even at large values of
the momentum transfer. Examples are the BSW model \cite{bsw,bsw03} and
the variants of the multiple diffraction model \cite{mblois,mcjp,mm,covoetal}.

Therefore, these phenomenological results are in agreement with
the conclusions of our empirical analysis, presented in Sect. \ref{sec:6.1}:
eikonal with zero gives rise to only one dip in the corresponding differential
cross section and a smooth decrease at large values of the momentum transfers.
In this sense, the model-independent features extracted from our analysis
corroborate the ingredients present in the BSW model and the variants of
the multiple diffraction model.

\vspace{0.2cm}

\noindent
$\bullet$ \textit{Quantitative aspects}

At this point it seems worthwhile to attempt to go further in the investigation 
on more quantitative
connections among model parametrizations with zero and our empirical results
for the imaginary part of the eikonal. We stress, however, the critical comment
at the beginning of Sect.~\ref{sec:6} on the limitation of our model independent results.

The idea is to generate a discrete set 
of points for the extracted
$\tilde\Omega(q)$, with the associated uncertainties from error
propagation, and compare with  model parametrizations presenting one zero. 
As we shall see, suitable
quantities for this comparison are, as before,
$q^8 \tilde\Omega(q)$
and also $|\tilde\Omega(q)|$.

\begin{figure}
\resizebox{0.48\textwidth}{!}{\includegraphics{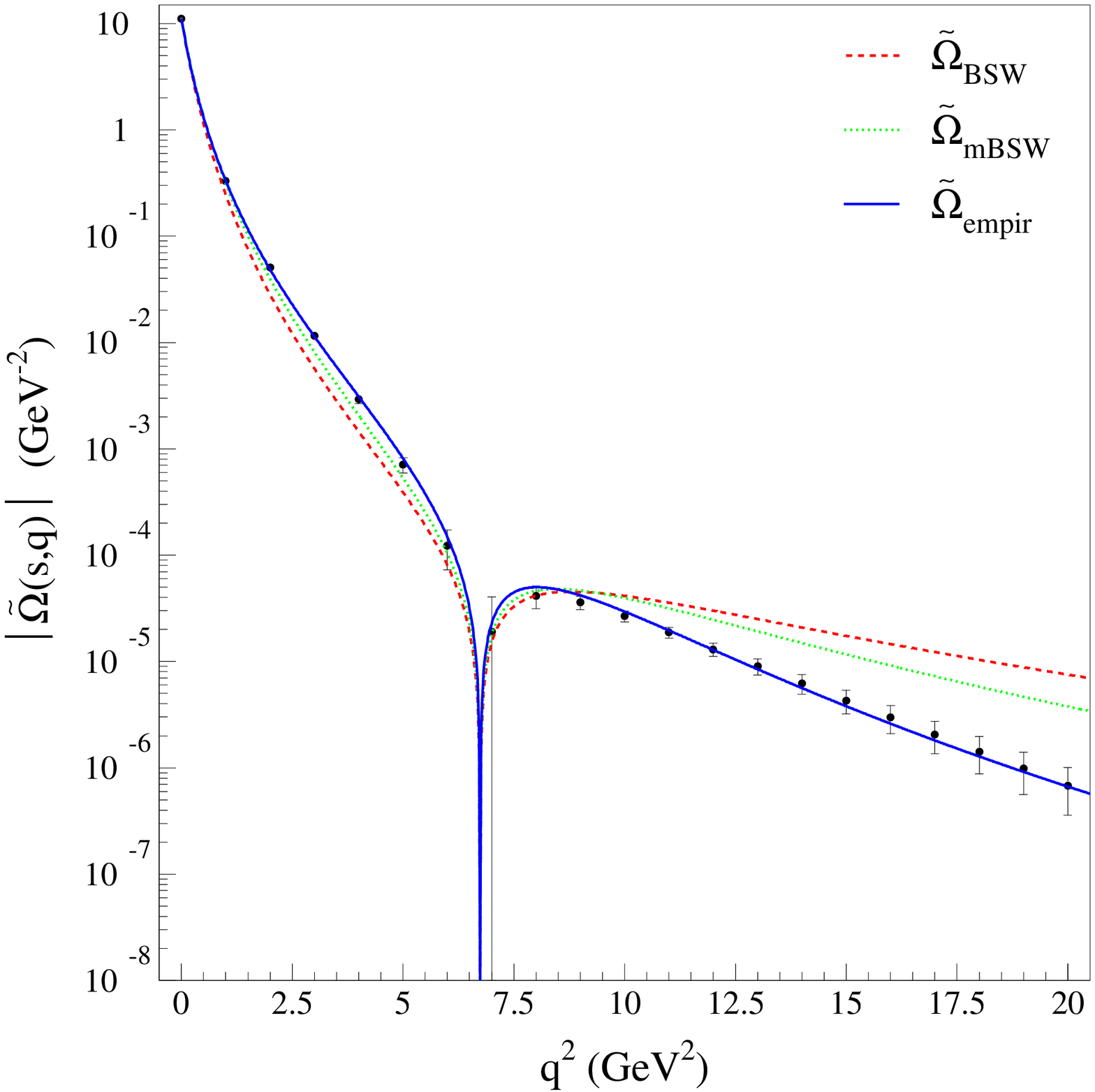}}
\resizebox{0.48\textwidth}{!}{\includegraphics{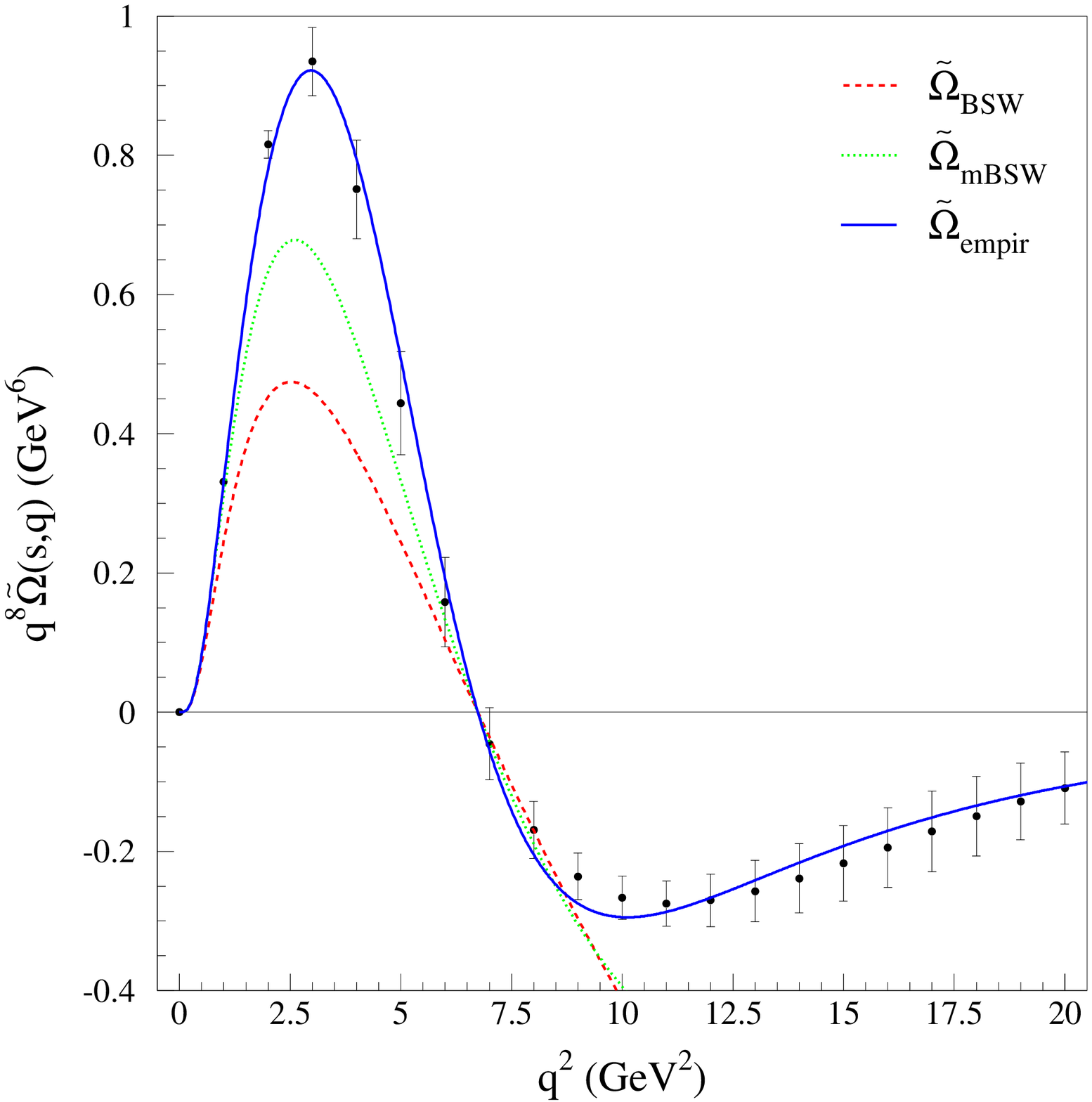}}
\caption{Generated points with uncertainties for the opacity, 
from the empirical fit to $pp$
scattering data at 52.8 GeV. The curves are the result of the data
reduction through $\tilde\Omega_{\mathrm{BSW}}(q)$, Eqs.
(\ref{eq:38}) and (\ref{eq:39}),  $\tilde\Omega_{\mathrm{mBSW}}(q)$, 
Eqs. (\ref{eq:38}) and (\ref{eq:40})
and $\tilde\Omega_{\mathrm{empir}}(q)$, Eq. (\ref{eq:44}).
(Table~\ref{tab:8}).}
\label{fig:14}
\end{figure}

In what follows we shall consider only the fit results obtained at 52.8 GeV, since the
original data set covers the largest region in momentum transfer
(up to 9.75 GeV$^2$), has one
of the largest number of points (adding data at 27.4 GeV) and the 
data reduction presented
a reasonable $\chi^2/DOF$ (Table~\ref{tab:5}). 
The empirical results for $pp$ scattering
at 52.8 GeV are displayed in Fig.~\ref{fig:14}  in the form of points with 
the propagated errors.

As regards models with eikonal zero, we consider the inputs of the BSW model for the
impact parameter dependence, Eqs. (\ref{eq:31}-\ref{eq:32}) and the original version of the
multiple diffraction model, Eqs. (\ref{eq:34}-\ref{eq:36}). For further 
discussion of these
two models, we introduce the
following notation for the imaginary part of the eikonal at \textit{fixed energy}:

\begin{eqnarray}
\tilde\Omega(q) =
\frac{C}{[1 + q^2/\alpha^2]^{2}[1 + q^2/\beta^2]^{2}} f(q),
\label{eq:38}
\end{eqnarray}
with either

\begin{eqnarray}
f(q) \rightarrow f_{\mathrm{BSW}} \equiv \frac{1 - q^2/q_0^2}{1 + q^2/q_0^2},
\label{eq:39}
\end{eqnarray}
or

\begin{eqnarray}
f(q) \rightarrow f_{\mathrm{mBSW}} \equiv \frac{1 - q^2/q_0^2}{1 + [q^2/q_0^2]^2},
\label{eq:40}
\end{eqnarray}
where the subscript mBSW stands for \textit{modified} BSW (referring to the 
square in the denominator).
This notation, introduced in \cite{mprd}, is useful,  since it allows 
for two distinct 
physical interpretations for the above eikonal, either in the Chou-Yang
or Glauber contexts:

1) a product of two form factors each one in the form introduced by BSW
(Chou-Yang context)

\begin{eqnarray}
G_1(q) =
\frac{1}{[1 + q^2/\alpha^2]}\frac{1}{[1 + q^2/\beta^2]} \sqrt{f(q)},
\label{eq:41}
\end{eqnarray}

2) a product of two form factors each one parametrized  as two simple poles 

\begin{eqnarray}
G_2(q) =
\frac{1}{[1 + q^2/\alpha^2]}\frac{1}{[1 + q^2/\beta^2]},
\label{eq:42}
\end{eqnarray}
by an elementary scattering amplitude (Glauber context)

\begin{eqnarray}
f_{\mathrm{BSW}}(q)
\quad \mathrm{or} \quad
f_{\mathrm{mBSW}}(q).
\label{eq:43}
\end{eqnarray}
The point is that, since $q_0^2$ represents the eikonal zero, in the former case
it is associated with the hadronic form factor and in the latter case
with the elementary amplitude.

For comparison with our empirical results at 52.8 GeV, we fix $q_0^2$ in the
above formulas to the 
extracted position of the zero at this energy, namely $6.74$ GeV$^2$ (this is also
the median of the values displayed in Table~\ref{tab:6}) and fit
the eikonals (\ref{eq:38}-\ref{eq:40}) to the generated points by means of the
CERN-Minuit code. The free parameters in both cases are $C$, $\alpha^2$
and $\beta^2$. The results are displayed in Table~\ref{tab:8} (2nd and 3rd columns)
and Fig.~\ref{fig:14}, with the following notation:

\begin{eqnarray}
\tilde\Omega_{\mathrm{BSW}}(q)
\quad \rightarrow \quad
\mathrm{Eqs.\ (\ref{eq:38})\ and\ (\ref{eq:39})}
\nonumber 
\end{eqnarray}

\begin{eqnarray}
\tilde\Omega_{\mathrm{mBSW}}(q)
\quad \rightarrow \quad
\mathrm{Eqs.\ (\ref{eq:38})\ and\ (\ref{eq:40})}
\nonumber 
\end{eqnarray}

\begin{table*}
\begin{center}
\caption{Results of the data reductions to the generated points in 
Fig.~\ref{fig:14}
($pp$ scattering at 52.8 GeV) through different parametrizations for the
imaginary part of the eikonal (see text).}
\label{tab:8}
\begin{tabular}{cccc}
\hline
\vspace{0.2cm}
 &  $\tilde\Omega_{\mathrm{BSW}}$  &  $\tilde\Omega_{\mathrm{mBSW}}$ & $\tilde\Omega_{\mathrm{empir}}$ \\
 & Eqs. (\ref{eq:38}) and (\ref{eq:39}) & Eqs. (\ref{eq:38}) and 
(\ref{eq:40}) & Eq. (\ref{eq:44})   \\
\hline
 $C$ (GeV$^{-2}$)      & 11.351 $\pm$  0.023  & 11.220 $\pm$ 0.039 & 11.155 $\pm$ 0.039 \\
 $\alpha^2$ (GeV$^2$)  & 0.704  $\pm$  0.014  & 0.746  $\pm$ 0.023 & 0.4534 $\pm$ 0.0093 \\
 $\beta^2$ (GeV$^2$)   & 0.704  $\pm$  0.015  & 0.746  $\pm$ 0.023 & 1.497  $\pm$ 0.047 \\
 $\chi^2/DOF$          & 207                  & 42                 & 0.50 \\
\hline
\end{tabular}
\end{center}
\end{table*}

We see that, although in both cases the modulus of the opacity 
is reasonably reproduced up
to $q^2 \sim$ 8 GeV$^2$, deviations occur above this region and as a 
consequence, the $\chi^2/DOF$ are too large (Table~\ref{tab:8}). 
Moreover, the plot of $q^8 \tilde\Omega(q)$
indicates that both parametrizations do not reach the generated points,
except near the fixed position of the zero and near the origin.

However, roughly, the result with $\tilde\Omega_{\mathrm{mBSW}}$ is nearer
the empirical points then that with the $\tilde\Omega_{\mathrm{BSW}}$. 
This effect is directly related to the square in the
$q^2/q_0^2$ term and may also explain the better reproduction of the
differential cross section data at \textit{large momentum transfers}
obtained with the multiple diffraction model (compare, for example,
the results for $\bar{p}p$ at 546 GeV in \cite{bsw03} and \cite{mcjp}).
Obviously, the position of the zero as obtained in both models,
from the phenomenological analysis, Eqs. (\ref{eq:33}) 
and (\ref{eq:37}), is not in agreement 
with the empirical result.

We have already stressed that our model independent analysis indicates only
one possible empirical result and that could explain some of the
differences with the para\-me\-tri\-zations discussed above. However,
even taking into account this limitation,
it may be useful, in the phenomenological context, to investigate what kind
of analytical parametrization can reproduce the generated points in 
Fig.~\ref{fig:14}
and that is our next task here.

\vspace{0.2cm}

\noindent
$\bullet$ \textit{Empirical parametrization for the eikonal}

We have tested several analytical parametrizations in order to
reproduce the extracted points in Fig.~\ref{fig:14}. The best result was obtained
with an additional square in the $q^2/q_0^2$ term present in the
denominator of the $f_{\mathrm{mBSW}}$ function, leading to the
following novel \textit{empirical} (empir) para\-me\-tri\-zation
for the opacity

\begin{eqnarray}
\tilde{\Omega}_{\mathrm{empir}}(q) =
\frac{C}{[1 + q^2/\alpha^2]^{2}[1 + q^2/\beta^2]^{2}} 
\frac{1 - q^2/q_0^2}{1 + [q^2/q_0^2]^4}.
\label{eq:44}
\end{eqnarray}

The results of the fit to the extracted points are displayed in 
Fig.~\ref{fig:14} and
Table~\ref{tab:8} (third colum), showing that the reproduction of the data is
quite good. Although this empirical parametrization may play some
important role in the phenomenological context, presently, we
can not provide a physical interpretation to it.
We note, however,  that a typical difference among all the
above parametrizations concerns the
asymptotic behavior, since for $q^2 \rightarrow \infty$ we have

\begin{eqnarray}
\tilde\Omega_{\mathrm{BSW}}(q) \sim
- \frac{1}{(q^2)^4}, 
\quad
\tilde\Omega_{\mathrm{mBSW}}(q) \sim - \frac{1}{(q^2)^5}, 
\nonumber
\end{eqnarray}

\begin{eqnarray}
\tilde\Omega_{\mathrm{empir}}(q) \sim
- \frac{1}{(q^2)^7}.
\nonumber
\end{eqnarray}
Some other aspects are discussed
in the following section.

\subsection{Electromagnetic and hadronic form factors}
\label{sec:6.3}

The Wu-Yang conjecture, associating the \textit{unknown} hadronic matter form 
factor
with electromagnetic form factor \cite{wy},
has played  a fundamental and historical role in the phenomenological
context. Also important, as we have recalled, has been the identification
of the hadronic form factor with the dipole parametrization,
Eq.~(\ref{eq:28}), for the Sachas electric form factor \cite{dl}.
These ideas date back to the end of the sixties and, on the other hand,
presently,
abundant data on the electromagnetic nucleon form factors are
available, at both
time-like ($q^2 < 0$) and space-like ($q^2 > 0$) regions, allowing
new insights in that old conjecture. Most important to our 
phenomenological purposes, is the
fact that recent experiments have indicated an unexpected decrease in the
proton electric form factor, as the momentum transfer increases, not
in disagreement with the possibility to reach zero just around
$q_0^2 \approx 7.5$ GeV$^2$. 

Therefore, to finish this work, it may be worthwhile
to explore some possible connections between
these results from the electromagnetic sector and those 
concerning the eikonal zero at $q_0^2 \approx$ 7 GeV$^2$, presented in the
preceding sections (some arguments on this respect have
already been discussed in \cite{cmm,mmm}). To this end we first summarize the 
new information on
the proton electric form factor and then discuss possible empirical
connections with our results. We shall not go into details, but only quote some results
of interest to our discussion. For recent detailed reviews on the subject
in both experimental and theoretical contexts, see, for example 
\cite{r1,r2}.

\subsubsection{Rosenbluth and polarization transfer results}

The traditional technique to experimentally investigate the nucleon
electromagnetic form factors has been the separation method by
Rosenbluth \cite{rose}, which is based on the measurement of the
differential cross section from unpolarized electron-nucleon
scattering. For the electron-proton case, the results have indicated a scaling
law for the ratio \cite{rose1,rose2}

\begin{eqnarray}
R_p = \mu_p \frac{G_E(q^2)}{G_M(q^2)} \approx 1,
\nonumber
\end{eqnarray}
where $\mu_p$ is the proton magnetic moment and $G_E$,  $G_M$ the
Sachas electric and magnetic form factors.

In 2000 - 2005, experiments with polarized electron beam, in polarization transfer
scattering,

\begin{eqnarray}
\vec{e}p\rightarrow e\vec{p},
\nonumber
\end{eqnarray}
have allowed simultaneous measurements 
to be made of the transverse and longitudinal
components of the recoil proton's polarization. By means of this
polarization transfer technique the ratio $G_E/G_M$ can be directly
determined with \textit{great reduction of the systematic uncertainties}
at large momentum transfers, $q^2: 4 - 9$ GeV$^2$. The surprising result
was the indication that this ratio decreases almost linearly with
increasing momentum transfers \cite{pt1,pt2,pt3}, leading even to a parametrization,
at large $q^2$, of the form \cite{pt2}
\begin{eqnarray}
R_p = 1 - 0.135( q^2 - 0.24),
\nonumber
\end{eqnarray}
which, by extrapolation, indicates a zero (change of signal) in $G_E$ at

\begin{eqnarray}
q_0^2 \approx 7.6\ \mathrm{GeV}^2.
\nonumber
\end{eqnarray}

From a theoretical point of view, radiative corrections associated 
with two-photon exchange process, have been extensively investigated
as possible source of the observed differences. As commented on before,
we shall not treat these aspects here; see \cite{r1,r2} for all the
details and references.

\subsubsection{The Proton Electric Form Factor}

\begin{figure}
\resizebox{0.48\textwidth}{!}{\includegraphics{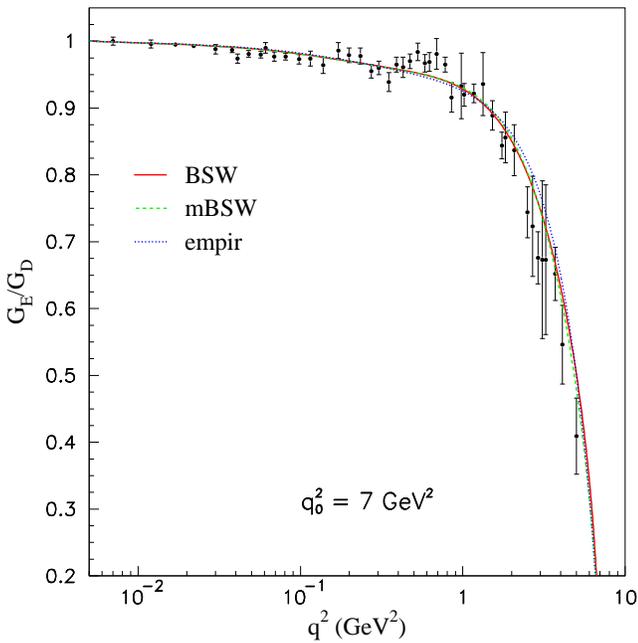}}
\caption{Experimental data on the ratio of the proton electric form factor to the
dipole parametrization, $G_E/G_D$ from \cite{ffdata} and fit results through 
BSW, mBSW and the empirical parametrizations, Eqs. (\ref{eq:45}), 
(\ref{eq:46}) and (\ref{eq:47}), respectively
and Eq. (\ref{eq:48}), with $q_0^2$ = 7 GeV$^2$ (Table~\ref{tab:9}).}
\label{fig:15}
\end{figure}

\begin{figure}
\resizebox{0.48\textwidth}{!}{\includegraphics{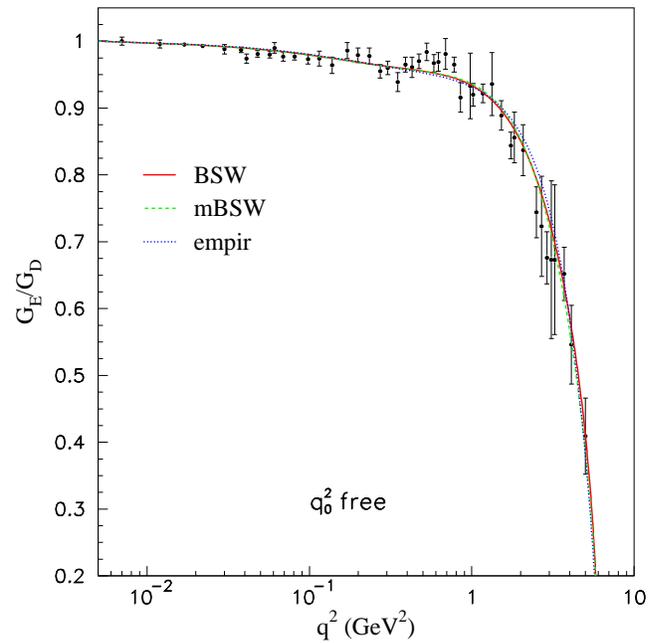}}
\caption{Same as Fig.~\ref{fig:15} with $q_0^2$ as a free fit parameter.}
\label{fig:16}
\end{figure}

Recently, a global analysis of the world's data on elastic electron-proton
scattering, taking into account the effects of two-photon exchange has
been performed. The analysis combines both 
the corrected Rosenbluth cross section and polarization transfer data, providing the
corrected values of $G_E$ and $G_M$ over the full $q^2$ range
with available data \cite{ffdata}. The results for the ratio 
$G_E/G_D$ between the proton
electric form factor and the dipole parametrization (with $\mu^2 = 0.71$ GeV$^2$),
covering the region $q^2 \approx 10^{-2} - 6$ GeV$^2$, are displayed in Figs.
\ref{fig:15} and \ref{fig:16}.

These data clearly show the deviation of $G_E$ from $G_D$ for
$q^2$ above $\approx 1$ GeV$^2$ and, from a strictly empirical
point of view, that $G_E$ might reach zero around $q_0^2 \approx$ 7 - 8 GeV$^2$.
Obviously, this zero might also be reached in an asymptotic process,
as predicted, for example, in the Unitary \& Analytic model \cite{dub}.

Anyway, in the context of the Wu-Yang conjecture, it may be worthwhile to compare the parametrizations for the
hadronic form factors from eikonal models with one zero (Sect. \ref{sec:6.2.4})
and the above data. To this end we return to the following parametrizations
for the hadronic proton form factor (Chou-Yang context), with the following notation

\begin{eqnarray}
G_{\mathrm{BSW}} =
\frac{1}{[1 + q^2/\alpha^2]}\frac{1}{[1 + q^2/\beta^2]} \sqrt{\frac{1 - q^2/q_0^2}{1 + q^2/q_0^2}},
\label{eq:45}
\end{eqnarray}

\begin{eqnarray}
G_{\mathrm{mBSW}} =
\frac{1}{[1 + q^2/\alpha^2]}\frac{1}{[1 + q^2/\beta^2]} \sqrt{\frac{1 - q^2/q_0^2}{1 + [q^2/q_0^2]^2}},
\label{eq:46}
\end{eqnarray}

\begin{eqnarray}
G_{\mathrm{empir}} =
\frac{1}{[1 + q^2/\alpha^2]}\frac{1}{[1 + q^2/\beta^2]} \sqrt{\frac{1 - q^2/q_0^2}{1 + [q^2/q_0^2]^4}}.
\label{eq:47}
\end{eqnarray}

The point is to
construct the ratio of each of the above formulas with the dipole
parametrization, Eq.~(\ref{eq:28}),

\begin{eqnarray}
\frac{G_i(q)}{G_D(q)}, \quad i = 
\mathrm{BSW},\ \mathrm{mBSW},\ \mathrm{empir}
\label{eq:48}
\end{eqnarray}
and perform the fits to the data in Fig. \ref{eq:15} through the code
CERN-minuit. In addition we consider two variants for the data
reduction:

\begin{description}

\item[\#1.] $q_0^2$ fixed to our average result in the ISR region,
 Eq.~(\ref{eq:21}), 
$q_0^2$ = 7.0 GeV$^2$ and $\alpha^2$ and $\beta^2$ as free
fit parameters.

\item[\#2.] $q_0^2$ as a free fit parameter together with 
$\alpha^2$ and $\beta^2$.

\end{description}

The results are shown in Figs. \ref{fig:15} and \ref{fig:16}, respectively
and 
the numerical results are displayed in Table~\ref{tab:9}.

\begin{table*}
\begin{center}
\caption{Results of the fit to the extracted ratio between the
proton electric form factor and dipole parametrization. All the parameters
in GeV$^2$.}
\label{tab:9}
\begin{tabular}{ccccc}
\hline
$q_0^2$ &      &  BSW    & mBSW     & empir        \\

        &       & Eq. (\ref{eq:45}) and (\ref{eq:48}) & 
Eq. (\ref{eq:46}) and (\ref{eq:48}) & Eq. (\ref{eq:47}) and (\ref{eq:48}) \\
\hline
& $\alpha^2$   & 1.550 $\pm$ 0.073 & 1.310 $\pm$ 0.064 &  1.156 $\pm$ 0.055 \\
7 GeV$^2$
& $\beta^2$    & 0.437 $\pm$ 0.010 & 0.446 $\pm$ 0.012 & 0.474 $\pm$ 0.014  \\
& $\chi^2/DOF$ & 1.36           & 1.34              &  1.79  \\
\hline
& $\alpha^2$    & 1.8068  $\pm$ 0.097  & 1.508 $\pm$ 0.084 &  1.328 $\pm$ 0.070 \\
free
& $\beta^2$     & 0.4192 $\pm$ 0.0090 & 0.423 $\pm$ 0.011 &  0.446 $\pm$ 0.012 \\
& $q_0^2$      & 6.06  $\pm$ 0.11   & 6.12  $\pm$ 0.13  & 6.04  $\pm$ 0.10 \\
& $\chi^2/DOF$ & 1.11                 & 1.11              &  1.41   \\
\hline
\end{tabular}
\end{center}
\end{table*}

We see that, although all the parametrizations provide good
visual descriptions of the data, the best statistical
results have been obtained with $G_{\mathrm{BSW}}$, Eq.~(\ref{eq:45}) and
$G_{\mathrm{mBSW}}$, Eq. (\ref{eq:46}) and $q_0^2$ as a free fit parameter: 
$\chi^2/DOF$ = 1.11 in both cases. Moreover, both fits indicate
$q_0^2 \approx$ 6.1 GeV$^2$, a value barely compatible with
our average estimation, Eq. (\ref{eq:21}).
The statistical results with $G_{\mathrm{empir}}$,
Eq.~(\ref{eq:47}), are not so good since 
the $\chi^2/DOF$ is higher.

These results suggest that parametrizations (\ref{eq:45}-\ref{eq:47}) have 
correlations with the recent global analysis on the proton electric form
factor \cite{ffdata}, a fact that may bring about new theoretical insights in the
phenomenological context. 
It seems to us that a striking aspect of the above empirical
results is the fact that they corroborate the old Wu-Yang conjecture
and just after a  complete change in the experimental knowledge
on the electromagnetic form factors along the 
years (Rosenbluth scaling versus polarization transfer results).

\section{Summary and final remarks}
\label{sec:7}

We have developed an empirical analysis of the differential cross
section data on elastic $pp$ scattering in the region
19.4 $\leq \sqrt s \leq$ 62.5 GeV. The analysis introduces two main improvements
if compared with a previous one \cite{cmm,cm}, the first associated
with the structure of the parametrization and the second with the selected
data ensemble.
We have also presented a critical discussion of the experimental data available,
checking, in some detail, that data at large momentum transfers
($q^2 >$ 3.5 GeV$^2$) do not depend on the energy in the particular region
23.5 $\leq \sqrt s \leq$ 62.5 GeV. Based on this information, we have included 
the data obtained at 27.4 GeV only in the 5 sets in the above
energy region and not at 19.4 GeV, as done in \cite{cmm}. With these 
improvements we have obtained better statistical results then in previous
analysis \cite{cmm,cm}.

As commented on at the beginning of Sect.~\ref{sec:6}, these fits represent only 
one solution. In fact, the data reduction of the differential cross sections
with $\sim$ 150 $DOF$ and $\sim$ 10 free parameters is a very complex
process and the main point is the lack of information on the contributions
from the real and imaginary parts of the amplitude beyond the
forward direction, leading to an infinity number of possible solutions. To our 
knowledge, the only model
independent information on the real part at $q^2 >$ 0 concerns a theorem by Martin,
which indicates a change of signal (zero) at small values of the momentum
transfer \cite{martin}. The exact position, however, can not be inferred.
In our approach, by including in the parametrization the experimental results
on $\sigma_{\mathrm{tot}}$ and $\rho$ at each energy, we correctly reproduce the
forward behavior in the region investigated. The zero in the real 
part is generated  by using two equal exponential contributions (in $q^2$)
in both real and imaginary parts ($m = 2$ and $n = 4, 5$ and $6$ in 
Eq.~(\ref{eq:11})).
However, the zero can also be generated without this constraint \cite{sma}.

Therefore, a general and detailed analysis on the physically acceptable
data reductions, constrained by model independent formal results, is
necessary and we are presently investigating the subject \cite{sma}.
Anyway, despite the above limitations in the present analysis, it allows
us to infer several novel qualitative and some quantitative results,
as summarized in what follows.

With the data reduction and by means of the semi-analytical method, 
the imaginary part of the eikonal (real opacity function), in the momentum transfer space, 
has been extracted, together
with uncertainty regions from error propagation. That was achieved within
approximation (\ref{eq:16}), justified by the fit results. Although 
the method provides model 
independent results for the eikonal in both $q$ and $b$ spaces, we
focused here only the question of the eikonal zero in $q$ space.
Different from the previous analysis \cite{cmm}, we obtained statistical
evidence for a change of sign in the imaginary part of the eikonal
only in the region 23.5 $ \leq \sqrt s \leq$ 62.5 GeV and not at
19.4 GeV. Moreover, the position of the zero in this energy
region is approximately constant with average value $q_0^2 =$ 7 $\pm$ 1
GeV$^2$, compatible with the result obtained in \cite{cm}, where
only the ISR data was considered.

The implication of the eikonal zero in the phenomenological context
has been also discussed in some detail. We have shown that models with
two dynamical contributions for the imaginary part of the
eikonal (positive at small and negative at large momentum transfers)
allows good descriptions of the differential cross section data in
the full $q^2$ region with available data. In this context the
BSW model play a central role due to both its theoretical basis
and the reproduction of the experimental data. We have also discussed
some analytical parametrizations for the extracted eikonal, either from
phenomenological models ($\Omega_{\mathrm{BSW}}$ and $\Omega_{\mathrm{mBSW}}$) or
by introducting a novel form
($\Omega_{\mathrm{empir}}$). 

Connections between the extracted eikonal and recent global analysis on the
proton electric form factor have also been discussed. In particular
we have shown that eikonal models presenting good descriptions of the elastic
hadron scattering, make use of effective form factors also compatible
with the proton electric form factor and in this case, the fits
have indicated a zero at $q_0^2 \approx$ 6.1 GeV$^2$. This compatibility between hadronic
and electric form factors
seems a remarkable fact if we consider all the
theoretical and experimental developments that took place after the original
conjecture by Wu and Yang.

We understand that all these empirical results can provide novel
and important insights in the phenomenological context, since,
through the Fourier transform, suitable inputs for the
``unknown" impact parameter contribution can be obtained. For example, in the case
of QCD-inspired models, the factorization in $s$ and $b$ at the
elementary level (Sect. \ref{sec:6.2.1}) allows, in principle, any
choice for the impact parameter contribution without losing the
semi-hard QCD connections ($\sigma_{gg}(s)$, for example). The use
of parametrizations (45-47) in the place of the dipole parametrization
at the elementary level ($qq$, $qg$, $gg$ contributions) may be much
more efficient in the description of the differential cross section
data at large values of the momentum transfer. That is, at least,
what our phenomenological analysis suggests.
We are presently investigating this subject.

Finally we would like to call attention to a central aspect related to the importance
of the differential cross section information at large momentum
transfers in any reliable model
independent analysis. Comparison of Figs. \ref{fig:9} and \ref{fig:10}
shows clearly that the lack of data at large momentum transfers
turns out to make very difficult or even impossible, any detailed knowledge of the
elastic scattering processes. Despite the technical difficulties in performing
experiments at a large momentum region, we think this should be an aspect to be
taken into account in the forthcoming experiments. We end this work
stressing once more the assertion by Kawasaki, Maehara and Yonezawa
\cite{kmy} ``Such experiments will give much more valuable information for the
diffraction interaction rather than to go to higher energies''.

\begin{acknowledgement}
       
We are thankful to FAPESP for financial support
(Contracts No.03/00228-0 and No.04/10619-9) and to A.F. Martini,
J. Montanha and E.G.S. Luna for discussions.

\end{acknowledgement}

\end{document}